\documentclass{aa}  

\usepackage{txfonts}
\usepackage{color, colortbl}
\usepackage[colorlinks, citecolor=blue, linkcolor=blue, urlcolor=blue]{hyperref}

\usepackage{graphicx}
\usepackage{txfonts}
\usepackage{pdflscape}	
\usepackage{color, colortbl}

\usepackage{newtxtext,newtxmath}

\usepackage[T1]{fontenc}
\usepackage{ae,aecompl}

\usepackage{pdflscape}	

\usepackage{listings}
\usepackage{pdflscape}
\usepackage{longtable}

\usepackage{graphicx}	
\usepackage{amsmath}	
\usepackage{amssymb}	
\usepackage{multirow}

\newcommand{\kms}{\,km\,s$^{-1}$}	
\newcommand{\gaia}{{\it Gaia}}
\newcommand{\stab}{54}
\newcommand{\rrlyr}{20}
\newcommand{\samp}{4247}

\usepackage{soul}

\begin{document}

\title{Milky Way archaeology using RR Lyrae and type II Cepheids I. The Orphan stream in 7D using RR Lyrae stars}

\author{Z. Prudil\inst{1} \and M. Hanke\inst{1} \and B. Lemasle\inst{1}
		\and
		J.~Crestani\inst{2,3,4} \and V.~F.~Braga\inst{3,5} \and M.~Fabrizio\inst{3,5}  \and A. J. Koch-Hansen\inst{1} G.~Bono\inst{2,3} \and E.~K.~Grebel\inst{1} 
		\and N.~Matsunaga\inst{6} \and M.~Marengo\inst{7} \and R.~da~Silva\inst{3,5} \and M.~Dall'Ora\inst{8} \and C.~E.~Mart\'inez-V\'azquez\inst{9} 
		\and G.~Altavilla\inst{3,5} \and H.~Lala\inst{1} \and
		B.~Chaboyer\inst{10} 
		\and I.~Ferraro\inst{3}
		\and G.~Fiorentino\inst{3} \and C.~Gilligan\inst{10}
		\and M.~Nonino\inst{11}
		\and F.~Th\'evenin\inst{12}}

\institute{Astronomisches Rechen-Institut, Zentrum f{\"u}r Astronomie der Universit{\"a}t Heidelberg, M{\"o}nchhofstr. 12-14, D-69120 Heidelberg, Germany \\ \href{mailto:prudilz@ari.uni-heidelberg.de}{prudilz@ari.uni-heidelberg.de} 
\and Dipartimento di Fisica, Universit\`a di Roma Tor Vergata, via della Ricerca Scientifica 1, 00133 Roma, Italy 
\and INAF -- Osservatorio Astronomico di Roma, via Frascati 33, 00078 Monte Porzio Catone, Italy
\and Departamento de Astronomia, Universidade Federal do Rio Grande do Sul, Av. Bento Gon\c{c}alves 6500, Porto Alegre 91501-970, Brazil
\and Space Science Data Center, via del Politecnico snc, 00133 Roma, Italy
\and Department of Astronomy, The University of Tokyo, 7-3-1 Hongo, Bunkyo-ku, Tokyo 113-0033, Japan
\and Department of Physics and Astronomy, Iowa State University, Ames, IA 50011, USA
\and INAF -- Osservatorio Astronomico di Capodimonte, Salita Moiariello 16, 80131 Napoli, Italy
\and Cerro Tololo Inter-American Observatory, NSF's National Optical-Infrared Astronomy Research Laboratory, Casilla 603, La Serena, Chile
\and Department of Physics and Astronomy, Dartmouth College, Hanover, NH 03755, USA
\and INAF -- Osservatorio Astronomico di Trieste, Via G.~B. Tiepolo 11, 34143 Trieste, Italy
\and Universit{\'e} de Nice Sophia-antipolis, CNRS, Observatoire de la C\^{o}te d’Azur, Laboratoire Lagrange, BP 4229, F-06304 Nice, France}

\date{\today}

\abstract
{We present a chemo-dynamical study of the Orphan stellar stream using a catalog of RR~Lyrae pulsating variable stars for which photometric, astrometric, and spectroscopic data are available. Employing low-resolution spectra from the Sloan Digital Sky Survey (SDSS), we determined line-of-sight velocities for individual exposures and derived the systemic velocities of the RR~Lyrae stars. In combination with the stars' spectroscopic metallicities and \gaia~EDR3 astrometry, we investigated the northern part of the Orphan stream. In our probabilistic approach, we found \rrlyr~single mode RR~Lyrae variables likely associated with the Orphan stream based on their positions, proper motions, and distances. The acquired sample permitted us to expand our search to nonvariable stars in the SDSS dataset, utilizing line-of-sight velocities determined by the SDSS. We found \stab~additional nonvariable stars linked to the Orphan stream. The metallicity distribution for the identified red giant branch stars and blue horizontal branch stars is, on average, $-2.13 \pm 0.05$\,dex and $-1.87 \pm 0.14$\,dex, with dispersions of $0.23$ and $0.43$\,dex, respectively. The metallicity distribution of the RR~Lyrae variables peaks at $-1.80 \pm 0.06$\,dex and a dispersion of $0.25$\,dex. Using the collected stellar sample, we investigated a possible link between the ultra-faint dwarf galaxy Grus\,II and the Orphan stream. Based on their kinematics, we found that both the stream RR~Lyrae and Grus\,II are on a prograde orbit with similar orbital properties, although the large uncertainties on the dynamical properties render an unambiguous claim of connection difficult. At the same time, the chemical analysis strongly weakens the connection between both. We argue that Grus\,II in combination with the Orphan stream would have to exhibit a strong inverse metallicity gradient, which to date has not been detected in any Local Group system.}

\titlerunning{The Orphan stream in 7D using RR Lyrae stars}
\maketitle


\section{Introduction}

The Milky Way (MW) halo holds fossil records of its formation history where passing smaller stellar systems were tidally disrupted by the Galactic gravitational field and subsequently mixed with the insitu MW stellar populations. The relics of past mergers can be found in the form of stellar streams and overdensities \citep[e.g.,][]{Helmi1999,Belokurov2006FS,Belokurov2007Orphan,Grillmair2006GD1,Grillmair2006Orphan,Bell2008,Newberg2016,Shipp2018,Malhan2018,Helmi2020}, with their spatial and kinematical distribution carrying an imprint of the underlying MW potential and mass distribution \citep[e.g.,][]{Johnston1999,Ibata2001,Newberg2002,Johnston2005,LawMaj2010Sag,Koposov2010,Kupper2015,Erkal2019}. The morphology of stellar streams may also provide insight into the dark matter subhalos predicted by the $\Lambda$ cold dark matter ($\Lambda$CDM) cosmology \citep[e.g.,][]{Dekel1986,Kauffmann1993,Springel2008}. In particular, dynamically cold streams can be utilized in the search for ''gaps'' \citep{deBoer2020} caused by a stream encounter with a dark matter subhalo \citep[e.g.,][]{Ibata2002,Carlberg2012,Erkal2015,Bonaca2019}, and they can possibly provide a lower limit on the size of dark matter subhalos \citep[e.g.,][]{Bode2001,Hu2000,Bullock2017}. Yet, a cautious treatment of the gaps is needed since epicyclic motion and giant molecular clouds can produce such stream features as well \citep{Amorisco2016,Ibata2020}.

The advent of large photometric, spectroscopic, and astrometric surveys uncovered a wealth of stellar substructures in the MW halo \citep[e.g.,][]{York2000,DES2018DR1,PS12010,GaiaEDR3Summary2020,Helmi2018Nature,Belokurov2018,Malhan2018}. Currently, the MW halo hosts over 60 known tidally disrupted remnants of globular clusters and dwarf galaxies \citep[e.g.,][]{Newberg2016,Mateu2018streams,Ibata2019}. Among the most prominent is the Orphan stellar stream, independently discovered by \citet{Grillmair2006Orphan} and \citet{Belokurov2007Orphan} in the Sloan Digital Sky Survey \citep[SDSS,][]{York2000}. 

The width of the Orphan stream ranges between $1-2$\,deg and spans across $210$\,deg on the sky \citep{Newberg2016,Koposov2019Orphan}, and it is traced out to a distance $\approx 60$\,kpc in both the southern and northern hemispheres \citep{Koposov2019Orphan}. The chemical composition of the likely stream members derived from SDSS low-resolution spectra exhibits a broad metallicity distribution with a mean at $-2.1$\,dex and spanning from $-1.5$\,dex up to approximately $-3.0$\,dex \citep{Newberg2010,Sesar2013Orphan}, both for blue horizontal branch (BHB) stars and for horizontal branch pulsators (RR~Lyrae stars, see below). The broad metallicity distribution (more than $1$\,dex) of the Orphan stream was later confirmed through low- and high-resolution spectroscopy \citep{Casey2013,Casey2014}, which solidified the dwarf-galaxy origin \citep{Sales2008} on the basis of their chemical abundance patterns. Also, such a broad metallicity distribution implies a prolonged star formation history, which is expected in the dwarf-galaxy paradigm.

The dwarf nature of the Orphan stream's progenitor is further hinted at in the stream's velocity dispersion $\sim 10$\kms \citep{Newberg2010}. A slightly lower velocity dispersion was reported by \citet[][$6.5$\kms]{Casey2013}, which was later corroborated by \citet{Koposov2019Orphan} and \citet{Fardal2019} placing the velocity dispersion at $\approx 5$\kms~and $\approx 7$\kms, respectively, still within the boundaries expected for a tidally disrupted, dwarf-like progenitor \citep[e.g.,][]{Gilmore2007,Koch2009,McConnachie2012}. The orbital modeling of the Orphan stellar stream suggests a prograde orbit with an eccentricity of $e\sim 0.7$, a pericentric distance of $16.4$\,kpc, and an apocentric distance of $90$\,kpc \citep{Newberg2010}. Recently, it has been shown that the velocity vector of the Orphan stream along its track is highly perturbed by the interaction with the Large Magellanic Cloud \citep{Erkal2019}.

The name Orphan comes from the long-standing issue of the unknown progenitor. Initial searches tried to link Orphan to the Ursa Major\,II and Segue\,1 dwarf spheroidal galaxies \citep{Fellhauer2007,Newberg2010}. Both dwarfs were later ruled out as Orphan progenitors on basis of their proper motions and distances \citep{Koposov2019Orphan} and satellite disruption modeling \citep{Sales2008}. One candidate remained, the ultra-faint dwarf (UFD) galaxy Grus\,II, found in the Dark Energy Survey \citep[DES,][]{Drlica2015,DES2018DR1}. Based on the sky position, proper motions, and distances Grus\,II, can be linked to the southern part of the Orphan stream \citep{Koposov2019Orphan}, although spectroscopic information such as line-of-sight velocities and chemical abundances are essential for solidifying their connection.

As a means of studying the Orphan stream, in our project we rely on pulsating variable stars of the RR~Lyrae class. RR~Lyrae variables are located inside the instability strip on the horizontal branch, and they are associated with old stellar populations with ages above 10\,Gyr \citep{Catelan2009,VandenBerg2013,Savino2020}. They are divided into three groups representing their pulsation mode: RRab (fundamental), RRc (first-overtone), and RRd (double-mode, pulsating simultaneously in the fundamental and first overtone mode) pulsators. Their pulsation periods are tightly connected to their luminosity \citep[on wavelengths redder than $R$-band, through period-luminosity-metallicity relations, PLZ,][]{Catelan2004,Muraveva2018,Neeley2019PLZ}, and thus RR~Lyrae stars serve as excellent distance indicators within the MW. In addition, the shape of their light curves reflects their chemical composition \citep{Jurcsik1996,Smolec2005,Hajdu2018}, thereby expanding their potential as tracers of the Galactic substructure and chemical composition. The aforementioned traits of RR~Lyrae stars made them invaluable in studies of stellar streams in the MW halo \citep[see, e.g.,][]{Sesar2013Orphan,Mateu2018streams,Hendel2018,Koposov2019Orphan,Price-Whelan2019}. In our work, we build on studies by \citet{Sesar2013Orphan}, \citet{Hendel2018}, \citet{Fardal2019}, and \citet{Koposov2019Orphan} who used RR~Lyrae stars to examine the Orphan stream.

We present the first paper of the series focused on the Milky Way archaeology using old classical pulsators. This paper aims at providing line-of-sight velocities and metallicities for the members of the Orphan stream alongside a discussion of a potential Orphan progenitor. The manuscript is organized in the following manner: Section~\ref{sec:DataSample} outlines the dataset we built together with the cuts we imposed and the distances that were estimated. Subsequently, in Section~\ref{sec:Method}, we describe the method we used for estimating the membership probability on basis of Bayesian inference. Section~\ref{sec:RR-streams} illustrates the spatial and kinematical distribution of RR~Lyrae variables from the assembled catalog associated with the Orphan stream. From the properties of the RR Lyrae population we were also able to recover non-pulsating stars in the SDSS catalog that are likely Orphan members. Both the method and the properties of these stars are described in Sections~\ref{sec:Method} and \ref{sec:RR-streams}. In Section~\ref{sec:diss} we discuss the possible metallicity gradient in the Orphan stream together with the orbital and chemical properties of Orphan members in context with the proposed Orphan progenitor. Final remarks are provided in Section~\ref{sec:Summary}.

\section{Properties of the RR Lyrae sample}  \label{sec:DataSample} 

As initial sample of RR Lyrae stars, we used the catalog of pulsating variables from the early second data release of the \gaia~mission \citep[DR2][]{Clementini2019} and found matches in the early third data release of the \gaia~source table \citep[EDR3,][]{GaiaEDR3Summary2020} in combination with RR~Lyrae stars identified in the Catalina sky survey \citep[CSS,][]{Drake2009} to avoid possible misclassification \citep{Molnar2018}. This sample provided us with some of the pulsation properties (pulsation periods) and astrometry \citep[precise coordinates and proper motions;][]{Lindegren2020GaiaAstrometry} necessary for our study. 

Subsequently, we cross-matched our RR~Lyrae sample with the spectroscopic part of the fifteenth data release of the SDSS \citep{Aguado2019}. The SDSS provides spectra collected over two decades using two multi-object fiber-fed spectrographs, namely SDSS\footnote{Used for the two phases of the Sloan Extension for Galactic Understanding and Exploration surveys \citep[SEGUE\,I and SEGUE\,II][]{Yanny2009SegueI,Eisenstein2011SegueII}.} and BOSS,\footnote{Designed for the Baryon Oscillation Spectroscopic Survey \citep{Smee2013,Dawson2013Boss}.} which share comparably low-resolutions ($R\sim 2000$) and a similar wavelength range from approximately $3600$\,\AA~to $10\,400$\,\AA. Both spectrographs use optical fibers that are plugged into the plates for a given observational field (640~fibers per plate for SDSS and 1000~fibers for BOSS plates), and have blue ($\approx$ 3600\,\AA~--~6000\,\AA) and red ($\approx$ 5800\,\AA~--~10\,400\,\AA) channels which are in the postprocessing co-added into the final data product \citep{Stoughton2002}. 

SDSS targeted stellar objects mainly in the range 14 - 20\,mag in the $g$-band, covering a large portion of the northern sky. Individual targets are given a \texttt{specObjID} identifier, which is generated based on the Modified Julian Date (MJD) of the observation (midpoint of the exposure), plate, and fiber ID. A fraction of our RR~Lyrae stars has been observed multiple times using different fibers, plates, and in some cases by both spectrographs. Each cross-matched RR~Lyrae star\footnote{Based on equatorial coordinates with a radius of 10 arcsec.} has one \texttt{bestObjID} identifier, which serves as a reference throughout our study, and one or several \texttt{specObjID}'s. We recovered spectroscopic data for the cross-matched sample from the SDSS Science Archive Server\footnote{\url{https://dr15.sdss.org/sas/dr15/}}. The retrieved data products contained the co-added (merged across epochs and for both channels) spectra together with the individual exposures for both channels (blue and red) and the precise time of the observation in MJD. The method for obtaining systemic velocities (corrected for the pulsation velocity) for individual RR~Lyrae variables is described in Appendix~\ref{ap:SpectraRad}. 

We note that the SDSS provides stellar parameters (e.g., metallicities, effective temperatures, and radial velocities) that were derived by the SEGUE stellar parameter pipeline \citep[SSPP,][]{Lee2008SsppI, Lee2008SsppII, Allende2008} for a large portion of our sample. These parameters were derived from the co-added spectra taken over several hours (sometimes across several days). Our targets rapidly change their radius \citep[with radial velocity amplitudes up to 130\kms,][]{Liu1991,Sesar2012} and effective temperatures $\approx$\,1000\,K \citep[e.g.,][]{For2011chem,Pancino2015,Jurcsik2018} in a matter of hours. Therefore, we used the combined spectra only for a comparison to our stellar parameters that were derived from the individual spectra (usualy taken with 900\,s exposures). 

To secure the purity of our sample, we obtained multi-epoch photometry from the CSS for our cross-matched \gaia~-~SDSS sample\footnote{Using the web interface: \url{http://nesssi.cacr.caltech.edu/cgi-bin/getmulticonedb_release2.cgi}.}. The CSS observes a portion of the northern and southern sky in the effort to find and monitor near-Earth objects, and as a by-product provides a large catalog of variable objects \citep{Drake2013,Drake2013stream,Drake2014CatVari,Abbas2014}. The CSS conducts unfiltered observations \citep[with a subsequent calibration to $V$-band using Landolt standard star catalog,][]{Landolt2007,Landolt2009} to increase the signal-to-noise ratio and detects faint objects down to $\sim$\,20\,mag with a single 30\,s exposure \citep{Drake2013}. The number of epochs for each object ranges from a few dozens to almost a thousand with an average uncertainty of 0.1\,mag. We verified the periodicity of the objects in our initial sample and obtained their ephemerides and pulsation properties. The details of this analysis can be found in Appendix~\ref{ap:Photometry}.

\subsection{The astrometric sample} \label{sub:AstroSamp}

For the purpose of using our catalog to study stellar streams, a precise astrometric solution including distances and a thorough treatment of their uncertainties is essential. In order to carefully assess the proper motions for individual variables we followed \citet{Hanke2020} and \citet{Prudil2020Disk}, and utilized the values provided by \textit{Gaia}'s EDR3 for proper motions in right ascension and declination ($\mu_{\alpha^{\ast}}$, $\mu_{\delta}$), their uncertainties ($\sigma_{\mu_{\alpha}^{\ast}},~\sigma_{\mu_{\delta}}$), covariances ($\rho_{\mu_{\alpha}^{\ast},~\mu_{\delta}}$), and re-normalized unit weight error (RUWE\footnote{The RUWE serves as an informative statistic on the quality of the astrometric five-parameter solution. We refer the interested reader to the technical note \url{http://www.rssd.esa.int/doc_fetch.php?id=3757412} for more details.}). 

In the first step, we scaled the covariance matrix, $\Sigma$, by the RUWE$^{2}$ factor, and diagonalized the resulting scaled covariance matrix by its eigenvectors (resulting in the transformed $\Sigma^{\ast}$). Using the eigenvectors of the covariance matrix, we transformed the vector composed of the stars' proper motions, \textbf{V}, and required at least 3$\sigma$ confidence in the scaled sum of the transformed proper motions:
\begin{equation} \label{eq:PMcut}
\sqrt{ \sum \mathbf{V}^{2} / \text{tr}(\mathbf{\Sigma^{\ast}}) } > 3.0 \\.
\end{equation}
This reduced our sample size from \samp~to 3970 RR Lyrae with at least 3$\sigma$ significant proper motions.

\subsection{Distance estimates} \label{sub:DistEst}

The connection of the RR Lyrae stars' pulsation periods, metallicities, and luminosities permits us to estimate a distance to a given RR~Lyrae star with an uncertainty on the order of three and ten percent for infrared and optical data, respectively \citep{Neeley2017}. The literature provides many PLZ relations both from the theoretical \citep[e.g.,][]{Catelan2004,Marconi2015,Marconi2018PLZ}, and observational studies \citep[e.g.,][]{Muraveva2018,Neeley2019PLZ}. The importance of metallicity in the PLZ relations and distance calculation is small as we move from the optical to the infrared wavelengths, it does not completely disappear, and the absence of a metallicity estimate for an individual star introduces an additional source of uncertainty on its distance estimate. 

Our data set is composed of unfiltered CSS photometry for which we estimated the mean magnitude based on a Fourier decomposition (see Appendix~\ref{ap:Photometry}). Unfortunately, absolute magnitudes of RR~Lyrae stars in the $V$-band are strongly dependent on metallicity, and not on pulsation period \citep[see][]{Catelan2004,Marconi2018PLZ,Muraveva2018}. 

To overcome this drawback, one needs to move from the visual wavelengths more toward the near-infrared or rely on the period-Wesenheit-metallicity (PWZ) relations, which provide a solid diagnostic for individual RR~Lyrae distances due to its low metallicity dependence. For this reason, we decided to cross-match our RR~Lyrae sample with the \text{PanSTARRS-1}\footnote{Panoramic Survey Telescope and Rapid Response System.} \citep[PS1,][]{Chambers2016} catalog of RR~Lyrae stars \citep{Sesar2017a}, and utilized their flux-averaged $i$-band magnitudes. The PLZ in the PS1 $i$-passband is strongly dependent on the pulsation period and only marginally on metallicity \citep[see table~1 in][]{Sesar2017a}. In order to estimate distances to the first-overtone pulsators we needed to transform their pulsation periods ($P_{\rm 1O}$ -- pulsation period of the first overtone mode) into the corresponding fundamental periods ($P_{\rm F}$ -- pulsation period of the fundamental mode) using the relation from \citet{Iben1971} and \citet{Braga2016}:
\begin{equation}
\text{log} P_{\rm F} = \text{log} P_{\rm 1O} + 0.127 \\.
\end{equation}
We note that there are several other approaches on how to transform the pulsation periods of RRc type stars \citep[e.g.,][]{DiCriscienzo2004,Coppola2015}, but their effect on the resulting absolute magnitude and subsequently distance is only marginal, and is completly covered by the total error budget of the absolute magnitude of a given star. To obtain metallicities for the $i$-band PLZ relation, we used samples analyzed by \citet{Fabrizio2019} and \citet{Crestani2020} which largely ($>$90\%) overlap our sample. To account for the missing metallicity in the remaining ten percent of the stars in our sample, we assumed a single value using the average and standard deviation by \citet[$\text{[Fe/H]}=-1.55 \pm 0.51$\,dex,][]{Crestani2020} for halo RR~Lyrae stars. To account for the reddening of the sample stars we utilized the extinction maps from \citet{Schlafly2011}.

\begin{figure*}
\includegraphics[width=2\columnwidth]{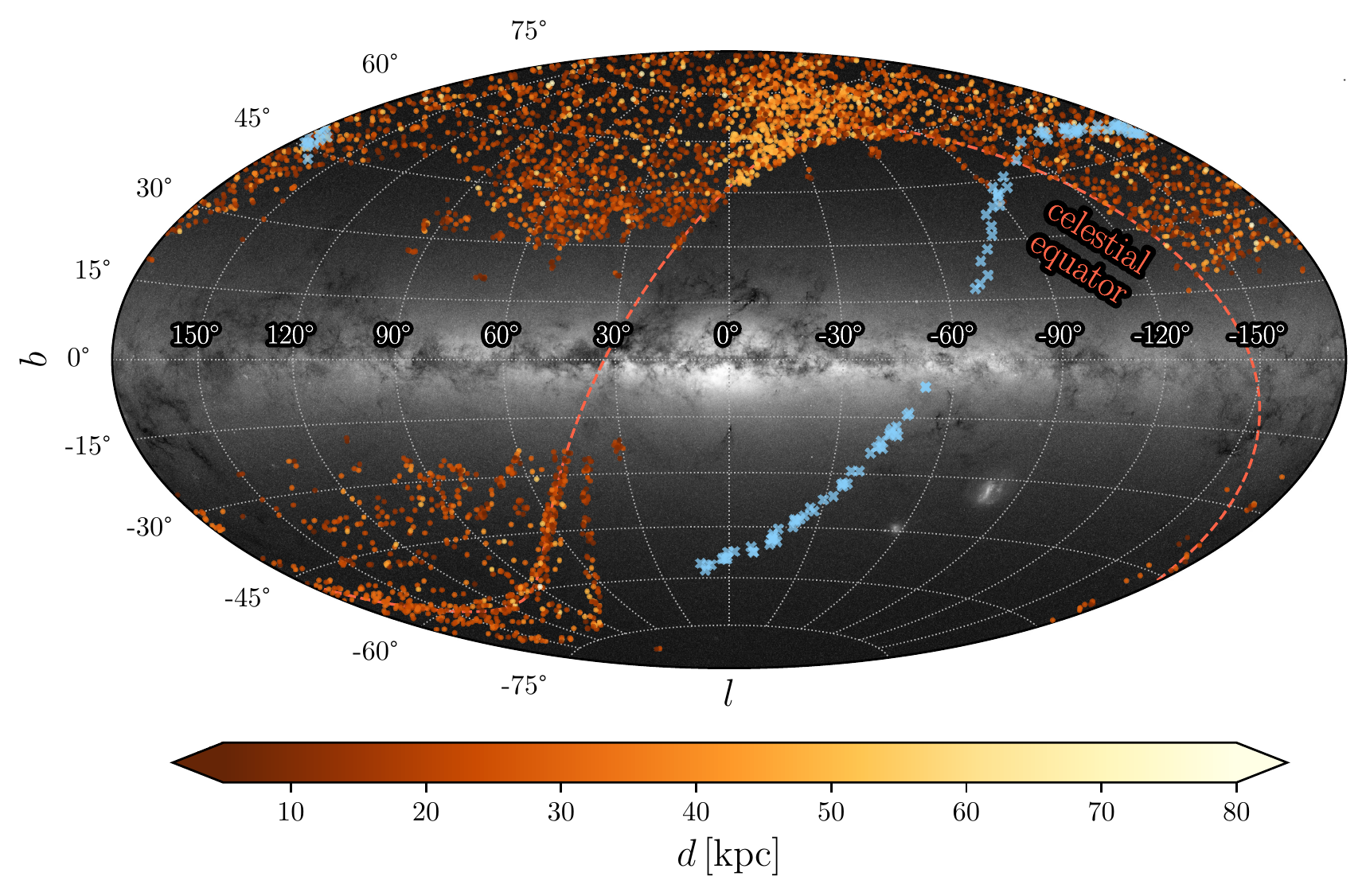}
\caption{Spatial distribution of RR~Lyrae stars (color-coded based on their distance) in Galactic coordinates. The light blue crosses denote the RR~Lyrae stars associated with the Orphan stream by \citet{Koposov2019Orphan}. \gaia's all-sky star density map is underlaid in the background as illustration. \textit{Image credit: Gaia Data Processing and Analysis Consortium (DPAC); A. Moitinho / A. F. Silva / M. Barros / C. Barata, University of Lisbon, Portugal; H. Savietto, Fork Research, Portugal.}}
\label{fig:MapOfSkyRRL}
\end{figure*}

To calculate distances, $d$, and their uncertainties, $\sigma_{d}$, we ran a Monte Carlo error analysis where we assumed a Gaussian distribution for the uncertainties on apparent magnitudes of $0.1$\,mag error on each $i$-band magnitude. We also varied the coefficients of the PLZ relation \citep[for the $i$-passband as listed in table~1 in][]{Sesar2017a}, within their errors, together with our assumed metallicities, reddening coefficients, and their associated uncertainties. The resulting distances range from $4$ to $100$\,kpc with the error budget varying from five to six percent. We note that our uncertainties are larger than generally reported for the PS1 survey of RR~Lyrae stars \citep[e.g.,][reported uncertainties around three percent]{Sesar2017a}. This is mainly due to our assumed error on the apparent magnitude, which we believe better represents the sparsity of PS1 observations. In Fig.~\ref{fig:MapOfSkyRRL} we depict the distribution of our selected RR~Lyrae variables with estimated distances. We show only the stars whose proper motions satisfy Eq.~\ref{eq:PMcut}.

As a validation check of our derived distances, we cross-matched our sample with the \textit{Spitzer} Merger History and Shape of the Galactic Halo (SMASH) sample of RR~Lyrae stars for the Orphan stream assembled by \citet{Hendel2018} and found $17$ variables in common. We detected a small offset of approximately $0.7$\,kpc between both sets of distances, a value roughly two to four times smaller than the individual uncertainties assigned to our distances and therefore negligible.

\section{Membership method} \label{sec:Method}

To assess a star's possible association with a given stellar stream, we employed a probabilistic approach similar to the one used for classical Cepheids in open clusters by \citet{Anderson2013}, and a study of MW globular cluster escapees in the halo \citep{Hanke2020}. In our analysis we establish membership probabilities based on the Bayesian framework that states that the posterior probability $p(A|B)$ of a model for the stream, $A$, and the data, $B$, is:
\begin{equation}\label{eq:BayesTheo}
p(A|B)  =  \frac{p(B|A) \times p(A)}{p(B)} \propto p(B|A) \times p(A) \\,
\end{equation}
which is a product of the likelihood function $p(B|A)$, our prior belief in an association, $p(A)$, and a normalizing constant, $p(B)$, representing the probability of observing the data \citep{Bayes1763}. Our analysis focused on connecting our sample of RR~Lyrae variables with the Orphan stellar stream which is sufficiently defined in equatorial coordinates $\alpha,\,\delta$, proper motions: $\mu_{\alpha^{\ast}},\,\mu_{\delta}$, and distances $d$.

Thus, we selected the prior to be a uniform probability distribution (with upper and lower boundaries) on the sky position $\alpha$:
\begin{equation}\label{eq:Prior}
p(A) = 1\;\texttt{if}\;\text{Min}\left (\left | \alpha^{\rm stream} - \alpha^{\rm RR \star} \right | \right ) < 5\,\text{deg}\;\texttt{else}\; 0 \\.
\end{equation}
For a simple description of stellar streams in a multi-parameter space, we used the Gaussian process (GP) regressor implemented in the \texttt{scikit-learn} library \citep{Pedregosa2012}. The GPs are a Bayesian nonparametric approach to regression, and they are a useful tool for nonlinear regression and classification. In the GP regressor we predict a continuous variable by specifying a suitable covariance function (kernel). In our case we selected the following set of kernels and their hyperparameters\footnote{We note that for the individual regressions we varied the individual covariance functions. The GP models for individual parameters will be provided at \url{https://github.com/ZdenekPrudil/Orphan2020}.}:
\begin{lstlisting}[basicstyle=\footnotesize, language=python, mathescape=true]
kernel = (ConstantKernel() +
          WhiteKernel(noise_level=2) +
          Matern(length_scale=2, nu=3/2))
          $\times$ 0.025$^2$ $\cdot$ DotProduct(sigma_0=1.0, 
                    sigma_0_bounds=(0.1, 10.0)).
\end{lstlisting}
The optimization of the kernels' hyperparameters is performed internally by the optimizer based on the maximization of the log marginal likelihood instead of the computationally expensive cross-validation. We refer the interested reader to \citet{Rasmussen2005} for a comprehensive and detailed description of GPs.

Using GPs, we fitted the parameters $\delta,\mu_{\alpha^{\ast}},\mu_{\delta}$, and $d$ as a function of $\alpha$ for the bona fide members of the Orphan stellar stream \citep{Koposov2019Orphan}, and obtained a GP regression model for the aforementioned parameters. The individual models, when provided with $\alpha$, predict values and covariances for a given parameter.

\begin{figure}
\includegraphics[width=\columnwidth]{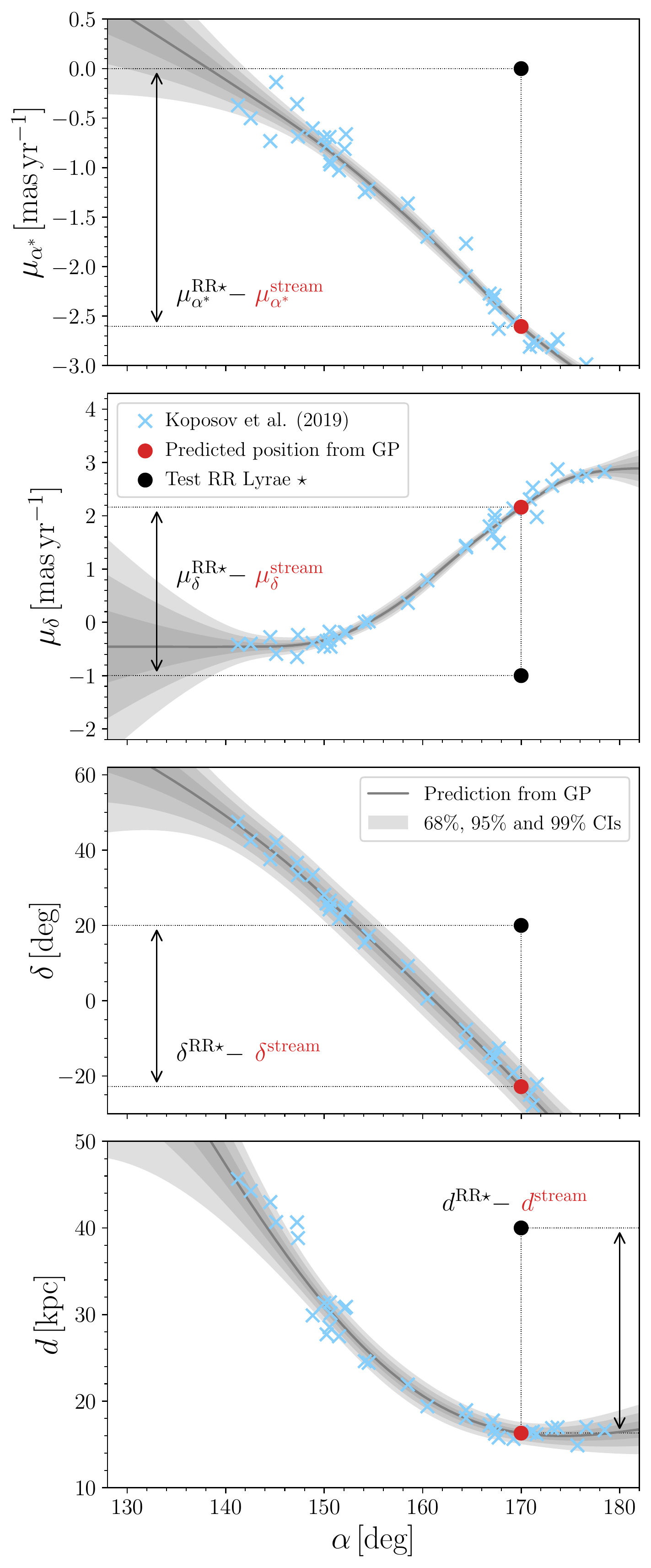}
\caption{Visual example of the membership analysis for the Orphan stellar stream using data from \citet[blue crosses,][]{Koposov2019Orphan}, with an artificially placed star (black dot), and its stream counterpart (red dot) at the same $\alpha$ and predicted values of $\delta, \mu_{\alpha^{\ast}},\mu_{\delta}$, and $d$. Black solid lines and gray shaded regions denote the GP regression and its confidence intervals (CIs, $\pm1,2,3\sigma$), respectively.}
\label{fig:Example}
\end{figure}

In order to estimate the conditional likelihood $p(B|A)$, we followed the example by \citet{Anderson2013} and \citet{Hanke2020}, and utilized the Mahalobis distance\footnote{Which in practice is a generalized Euclidean distance (with the identity covariance matrix), and is often used for the identification of outliers \citep{Myung2000}.} \citep{Mahalanobis1936Dist}:
\begin{equation}\label{eq:Mahal}
D_{M}^{2} = \left( \mathbf{x}^{\rm RR \star} - \mathbf{x}^{\rm stream} \right)^{\rm T} \mathbf{\Sigma}^{-1} \left(\mathbf{x}^{\rm RR \star} - \mathbf{x}^{\rm stream} \right)  \\,
\end{equation}
where $\mathbf{x}^{\rm RR \star}$ is a four-component vector composed of equatorial coordinates, proper motions, and distances ($\mathbf{x}^{\rm RR \star} = \left \{ \delta,\mu_{\alpha^{\ast}},\mu_{\delta},d \right \}$) for a given $\alpha$-coordinate. For obtaining a star's stream vector $\mathbf{x}^{\rm stream}$ we used as an input to the GP regression the star's equatorial $\alpha$ coordinate. The Gaussian regression models in turn yield a prediction for $\mathbf{x}^{\rm stream}$ and their variance for the streams' covariance matices. The visual depiction of our analysis can be found in Fig.~\ref{fig:Example}. $\mathbf{\Sigma}^{-1}$ represents the inverse sum of covariance matrices between an RR~Lyrae variable and a given stellar stream scaled by the squared RUWE. The covariance matrix for RR~Lyrae stars in our sample was constructed using the variances and correlation coefficients from \gaia~EDR3. Since our distances came from an independent source, we set their correlations with other parameters to zero. The stream covariance matrix is built using the prediction on the individual parameter from the GP regressor and only contains diagonal entries. To ensure that our stream quantities are independent of the variable sample (no covariance between $\mathbf{x}^{\rm RR \star}$ and $\mathbf{x}^{\rm stream}$) we removed cross-matched RR~Lyrae stars from the parent population of the stream sample for the GP regression of the stream distributions. 

Because of the assumption of a multivariate-normal error distribution the resulting $D_{M}^{2}$ is chi-squared distributed, in our case with four degrees of freedom (coordinate, proper motions, and distance). The likelihood function $p(B|A)$ can then be expressed as a $p$-value ($p_{\text{val}}$) of the $D_{M}^{2}$;
\begin{equation} \label{eq:PBA}
p(B|A) = 1 - p_{\text{val}}(D_{M}^{2}) \\. 
\end{equation}
The $p$-value is a probability metric for evaluating the null hypothesis, which in our case is a hypothesis test whether a star is or is not associated with a given stellar stream. A high $p$-value in Eq.~\ref{eq:PBA} highlights stars that we considered as outliers from the stream. Thus, our probability calculation mainly tags the stream's outliers (nonmembers). Conversely, if a high number of explored dimensions is provided, with strong constraints on the significance of individual parameters, then the probability of a star's membership in a given stream increases. We note that just as in any general case, the null hypothesis cannot be proven but only excluded. Thus, we treat the identified members as likely associations.

With the goal to distinguish between outliers and possible members, we selected for $p(A|B)$ a critical threshold of 0.05. Thus the RR~Lyrae stars in our sample with a higher $p(A|B)$ will be treated as tentative stream members.

\section{RR Lyrae and non-pulsating stars in the Orphan stream} \label{sec:RR-streams}

\subsection{RR Lyrae stars in the Orphan stream} \label{subsec:RRlyr}

Since its discovery \citep{Grillmair2006Orphan,Belokurov2007Orphan}, the Orphan stream has been targeted by various studies that provided several lists of possible candidates representing a variety of stellar types \citep[e.g., F-turnoff stars, BHB stars, RR~Lyrae stars, and K-giants,][]{Newberg2010,Sesar2013Orphan,Koposov2019Orphan,Casey2013}. The sample from \citet{Newberg2010} is based on the SDSS photometric and spectroscopic products, providing important spatial, dynamical, and chemical information about the Orphan stream, especially the metallicities of the BHB stars ([Fe/H]$=-2.1$\,dex), and their spread hint toward the progenitor of the Orphan stream being a dwarf galaxy.

The work by \citet{Sesar2013Orphan} confirmed the mean metallicity of the Orphan stream and its large spread found by \citet{Newberg2010}, and provided precise distances to individual RR~Lyrae stars effectively tracing the Orphan stream out to $55$\,kpc. The first detailed chemical abundance study of the Orphan stream by \citet{Casey2013} provided stream candidates based on their spatial, kinematic, and chemical properties. The associated K-giants exhibit a slightly more metal-rich composition ($\text{[Fe/H]} = -1.63$\,dex) than the BHB stars. We note that in the high-resolution spectroscopic study of \citet{Casey2014}, three high-probable candidates that can be kinematically and astrometrically associated with the Orphan stream exhibit a slightly lower average metallicity [Fe/H]$ = -2.01$\,dex.

We use the latest sample of possible stream members from the work by \citep[][and from here on we refer to it as the K19 reference sample]{Koposov2019Orphan}. The K19 sample includes \gaia~EDR3 and variable stars identification from \citet{Clementini2019}. It consists of 109 RR~Lyrae stars (106 fulfilling the condition in Eq.~\ref{eq:PMcut}) associated with the Orphan stream based on their spatial and kinematical properties. The Orphan reference sample spans both Galactic hemispheres, with a total coverage of about 210 degrees, and distances ranging from $\approx$ 10\,kpc to 60\,kpc. 

Our dataset relies on \gaia~EDR3 astrometric products and mainly on the \gaia~identification of RR~Lyrae stars \citep{Clementini2019} verified using the CSS and PS1 surveys, and covers primarily the northern Galactic hemisphere due to the SDSS footprint (see Fig.~\ref{fig:MapOfSkyRRL}). Our dataset offers a re-evaluated RR~Lyrae classification, improved distance estimates, metallicities, and systemic velocities for individual RR~Lyrae stars. The RR~Lyrae stars from the reference sample only served as an input for our membership analysis described in the previous section. From the K19 sample, 20 RR~Lyrae stars overlap with our dataset. The K19 sample does not contain uncertainties on individual distance estimates, which are based on visual magnitudes of individual RR~Lyrae variables, thus we assumed a general uncertainty of 10\,\% on the distance estimate for the Gaussian process regression. 

\begin{figure*}
\includegraphics[width=2\columnwidth]{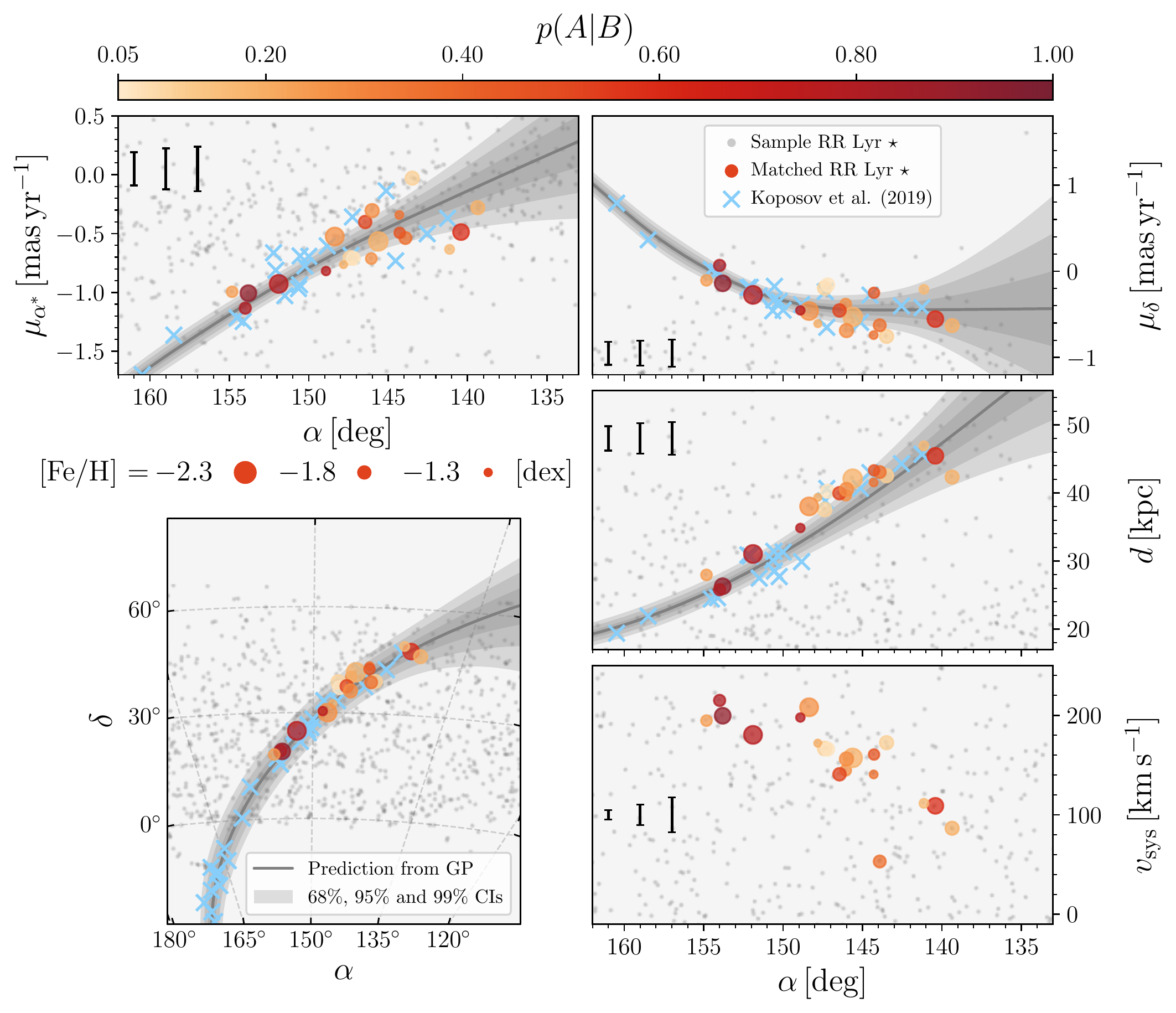}
\caption{Four-parameter association with the Orphan stream defined by the sample of RR~Lyrae stars from \citet[][denoted by light blue crosses]{Koposov2019Orphan} based on the spatial and astrometric properties of the studied sample. The RR~Lyrae stars associated with the Orphan stream (with a lower significance threshold set at 0.05) are color-coded based on the conditional probability $p(A|B)$. Gray dots represent rejected RR~Lyrae stars from our sample. Black lines and gray regions denote the GP fit to the reference sample and confidence intervals of a given interpolation, respectively. The metallicity of each star associated with the Orphan stream is depicted by varying its point size. The error bars in the left corners represent the $15.9$, $50$, and $84.1$ percentiles of individual parameter uncertainties for the RR~Lyrae variables linked with the Orphan stream.}
\label{fig:OrphanMem}
\end{figure*}

In Figure~\ref{fig:OrphanMem}, we show the results of our analysis for our sample of RR~Lyrae located in the vicinity of the K19 dataset. In our investigation, we identified \rrlyr~RR~Lyrae variables (13 RRab and 7 RRc-type pulsators) to be associated with the Orphan stream based on their equatorial coordinates, proper motions, and distances. From these stream associates, we recover 12 variables already present in the K19 sample. The remaining eight RR~Lyrae pulsators consist of three variables that were identified as members of the Orphan stream by \citet{Sesar2013Orphan} and \citet{Hendel2018}, while five are new discoveries. The likelihoods of stars not included in the K19 sample range from $p(A|B)=0.05$ (by construction owing to the adopted lower threshold) up to almost $p(A|B)=0.8$, with only four below $p(A|B)<0.2$. Similar to the K19 sample, we trace the Orphan stream from approximately $25$\,kpc to $47$\,kpc in distance across $32$\,deg on the sky. The proper motion ranges are $\mu_{\alpha^{\ast}} \approx \left(-1.13; -0.03 \right)$\,mas\,yr$^{-1}$ and $\mu_{\delta} \approx \left(-0.75; 0.07 \right)$\,mas\,yr$^{-1}$ and follow by construction the ranges of the K19 RR~Lyrae stars. Based on the likely stream members, the projected width of Orphan stream varies around $1-2$ deg, which is similar to the findings of \citet{Grillmair2006Orphan} and \citet{Belokurov2007Orphan}. We also report a higher average metallicity for the Orphan RR~Lyrae stars of [Fe/H]$=-1.80(6)$\,dex with a dispersion of $0.25$\,dex. This is significantly more metal-rich than previously reported by \citet[][average metallicity equal to $-2.1$\,dex]{Sesar2013Orphan}. This point will be discussed in Sect.~\ref{sec:stableOrphan}. In Fig.~\ref{fig:OrphanMem} we notice that one of the apparently associated RR~Lyrae variables does not fit the general systemic velocity trend. Thus, we consider it as a nonmember and remove it in the further analysis, whilst marking it with an asterisk in Table~\ref{tab:Orphan-RR}. The remaining $19$ RR~Lyrae stars were used to assess our systemic velocities with respect to the \texttt{RV\_ADOP} determined by the SSPP pipeline. Expectedly, we found a lower dispersion in our systemic velocities in comparsion to dispersion in \texttt{RV\_ADOP}, $11.0$\kms and $19.5$\kms, respectively.

Using the calculated distances and estimated systemic velocities, we specifically looked for RR~Lyrae stars beyond 50\,kpc \citep[the estimated apogalacticon of $90$~kpc by][]{Newberg2010}, and we found no RR~Lyrae stars in our sample that could be considered as a continuation of the Orphan stream. As an additional corroboration of our Orphan RR~Lyrae candidates, we looked at their distribution in the period-amplitude plane and searched for high-amplitude short-period RR~Lyrae variables \citep[HASP,][]{Fiorentino2015}. The HASP RR~Lyrae stars are characterized by short pulsation periods ($P < 0.48$\,day) and high amplitudes (in $V$-band above $0.75$\,mag). They often occur in systems with high metallicity \citep[higher than $-1.5$\,dex, such as the Galactic bulge, metal-rich globular clusters, and partially also in the Galactic halo,][]{Fiorentino2015}. Based on Orphan's low metallicity we would not expect HASPs to be found in the Orphan stellar stream and we note that indeed none of our Orphan associated RR~Lyrae stars belong to the HASP group. Although one HASP RR~Lyrae star has been identified in the southern portion of the Orphan stream by \citet{MartinezVazquez2019} which is probably caused by the large dispersion in the metallicity distribution of Orphan RR~Lyrae stars that covers regions with [Fe/H]$>-1.5$\,dex and permits such possibility.

\subsection{Nonvariable stars in Orphan} \label{sec:stableOrphan}

Building upon the approach for RR~Lyrae stars, we performed a similar analysis with the remaining stellar sample of the SDSS. To this extent, we searched for objects analyzed by the SSPP pipeline, restricting the sample to those objects with determined $T_{\rm eff}$. Utilizing SSPP products, we obtained their atmospheric parameters ($T_{\rm eff}$, log\,$g$, [Fe/H]) together with their heliocentric line-of-sight velocities. The nonvariable sample, as we refer to it, was subsequently cross-matched using equatorial coordinates with the \gaia~EDR3 catalog to acquire their proper motions and photometric properties ($G$, $G_{\rm BP}$, and $G_{\rm RP}$ magnitudes). Regarding the proper motion significance, we required the same significance as in the case of the RR~Lyrae sample to remove possible outliers.

For our nonvariable sample, we proceeded with our method outlined in Sect.~\ref{sec:Method} (using our identified sample of Orphan RR~Lyrae stars as the parent population) with two differences. Firstly, instead of using spectrophotometric distances, which can be prone to many systematics, we substituted the distance in the $\mathbf{x}^{\rm \star}$ vector with the systemic velocity $\left( \mathbf{x}^{\rm \star} = \left \{ \delta,\mu_{\alpha^{\ast}},\mu_{\delta}, v_{\rm sys} \right \} \right) $, thus slightly favoring the kinematical over the spatial association. Secondly, we only looked for tentative members close to the stream itself, thus narrowing our uniform flat prior from five degrees to one degree. As an additional criterion, we adopted cuts on metallicities and log\,$g$ to select stars above the main sequence and thus remove the majority of the contributions from the Galactic disk:
\begin{equation} \label{eq:AddCoef}
\text{[Fe/H]} < -1.0\text{\,dex} \\ \cap \\
\text{log}\,g < 4.0\text{\,dex} \\.
\end{equation}
Following this approach, we recovered \stab~nonvariable stars likely associated with the Orphan stream as traced by our sample of RR~Lyrae variables (listed in Table~\ref{tab:Orphan-Stab}). We also recovered four stars that were previously identified as RR~Lyrae stars in the \gaia~DR2 and PS1 surveys. Using CSS photometry, we were able to classify three of them as double-mode RR~Lyrae pulsators. The one remaining variable has an uncertain classification. All four stars did not enter our initial analysis of single-mode RR~Lyrae stars and are denoted with an asterisk in Table~\ref{tab:Orphan-Stab}. The distributions of astrometric and kinematical parameters of the associated nonvariables are depicted in Fig.~\ref{fig:Orphan-stable}. 

\begin{figure}
\includegraphics[width=\columnwidth]{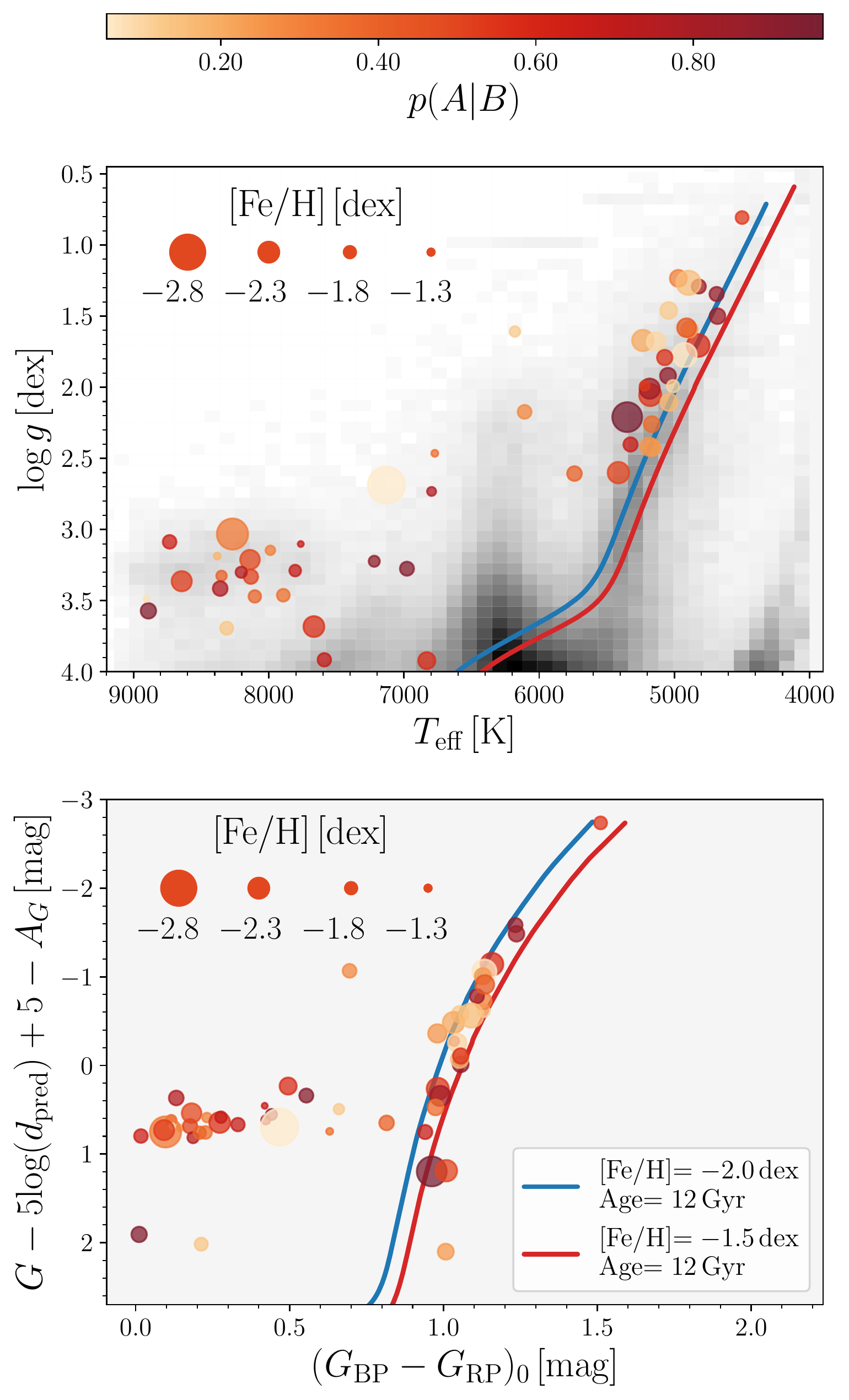}
\caption{Kiel diagram (top panel, using quantities from the SSPP pipeline), and the color-magnitude diagram (bottom panel) for stars likely associated with the Orphan stellar stream color-coded based on the probability and with varying point size denoting the metallicity. The blue and red lines represent isochrones from the MESA Isochrones \& Stellar Tracks \citep[MIST,][]{Dotter2016,Choi2016,Paxton2011,Paxton2013,Paxton2015} database for two different metallicities. The gray contours in the top panel represent the entire star matched sample of SDSS-\gaia.}
\label{fig:OrphanStabCMD}
\end{figure}

Utilizing the spectroscopic products (surface gravities and effective temperatures) determined by the SSPP pipeline and the dereddened photometry from \gaia~EDR3, we constructed the Kiel diagram (log\,$g$ vs. $T_{\rm eff}$) and the color-magnitude diagram for stable stars associated with the Orphan stream (see Fig.~\ref{fig:OrphanStabCMD}). To deredden \gaia~apparent magnitudes, we used the extinction coefficients from \citet[][see their table~2]{Casagrande2018} in combination with the dust maps derived by \citet{Schlafly2011}. The $G$ magnitudes of each stable star were corrected by the distance modulus estimated from the Gaussian process regression of our RR~Lyrae sample given its right ascension. 

In the top panel of Fig.~\ref{fig:OrphanStabCMD} we clearly identify the red giant branch (RGB, defined as $T_{\rm eff} < 5500$\,K and log\,$g < 3$\,dex, seen in Fig.~\ref{fig:OrphanStabCMD}) with several stars possessing a high membership probability ($p(A|B) > 0.5$). In addition, also BHB stars between $8000$ and $9000$\,K, and log\,$g$ ranging from $3.0$ to $3.5$\,dex were observed. We notice a discrepancy between the upper and lower panels, where for the upper panel (built with the SDSS spectroscopic products) an isochrone of metallicity $-2.0$\,dex provides a good fit, in contrast to \gaia~data where an isochrone of higher metallicity ($-1.5$\,dex) is necessary. We believe that this inconsistency is rooted in the stellar parameters derived by the SDSS: figure~A2 in \citet{Smolinski2011sspp} shows trends between stellar parameters $T_{\text{eff}}$, log\,$g$, and [Fe/H] derived by the SDSS and those from high-resolution studies. Similar trends in stellar parameters of the SDSS survey were also independently reported by \citet{Hanke2018} and \citet[][based on monometallic globular clusters]{Hanke2020}. 

\section{Discussion} \label{sec:diss}

The full 7D\footnote{Equatorial coordinates, distances, proper motions, line-of-sight velocities, and metallicities.} chemo-dynamical distribution of RR Lyrae stars likely associated with the Orphan stream permits us to examine their orbital parameters with respect to an assumed static MW potential. Jointly with chemical information in the form of [Fe/H] for nonvariable stars associated with the Orphan stream (see Sect.~\ref{sec:stableOrphan}) we can search for its possible progenitor. We focus on comparing with the work by K19, who provided a detailed examination of the properties of a possible Orphan progenitor regarding the stream RR~Lyrae population. K19 also discussed likely progenitors among several globular clusters and dwarf galaxies based on the spatial ($\alpha$, $\delta$, and distances), and proper motion spaces.  

\subsection{On a possible metallicity gradient in Orphan}

The metallicity of RGB and BHB stars centers at $-2.13 \pm 0.05$\,dex, and $-1.87 \pm 0.14$\,dex, with dispersions of $0.23$ and $0.43$, respectively. The average values are in good agreement with previous studies by \citet{Newberg2010} and \citet{Sesar2013Orphan}, who find an average metallicity of $-2.1$\,dex among RR~Lyrae stars associated with the Orphan stream. \citet{Sesar2013Orphan} also reported a metallicity gradient in their sample of RR~Lyrae stars. We explored this possibility by first cleaning the sample based on the \gaia~astrometry, following the same steps as in the case of our RR~Lyrae sample. From a total of $50$ RR~Lyrae stars in the \citet{Sesar2013Orphan} catalog we recovered $20$ likely members of the Orphan stream. Following \citet{Sesar2013Orphan} we calculated the Kendall's $\tau$ coefficient\footnote{The Kendall's correlation coefficient, $\tau$, is a nonparametric correlation test, thus independent of any assumptions on the distribution of the tested samples.} \citep{KendalTest1938} for the stream longitude, $\phi_{\rm 1}$  (calculated through the coordinates tranformation matrix from K19), with respect to the metallicity for these \rrlyr~single mode RR~Lyrae stars that are likely Orphan members, and we obtained $\tau_{\phi_{\rm 1}}^{\rm [Fe/H]} = -0.41 \pm 0.11$. This is very similar to the value reported by \citet{Sesar2013Orphan} and also significant\footnote{We note that we calculated the uncertainty on $\tau_{\phi_{\rm 1}}^{\rm [Fe/H]}$ through a Monte Carlo error simulation where we assumed a Gaussian distribution for errors on the metallicity ($\sigma_{\text{[Fe/H]}} = 0.2$\,dex).}.

\begin{figure}
\includegraphics[width=\columnwidth]{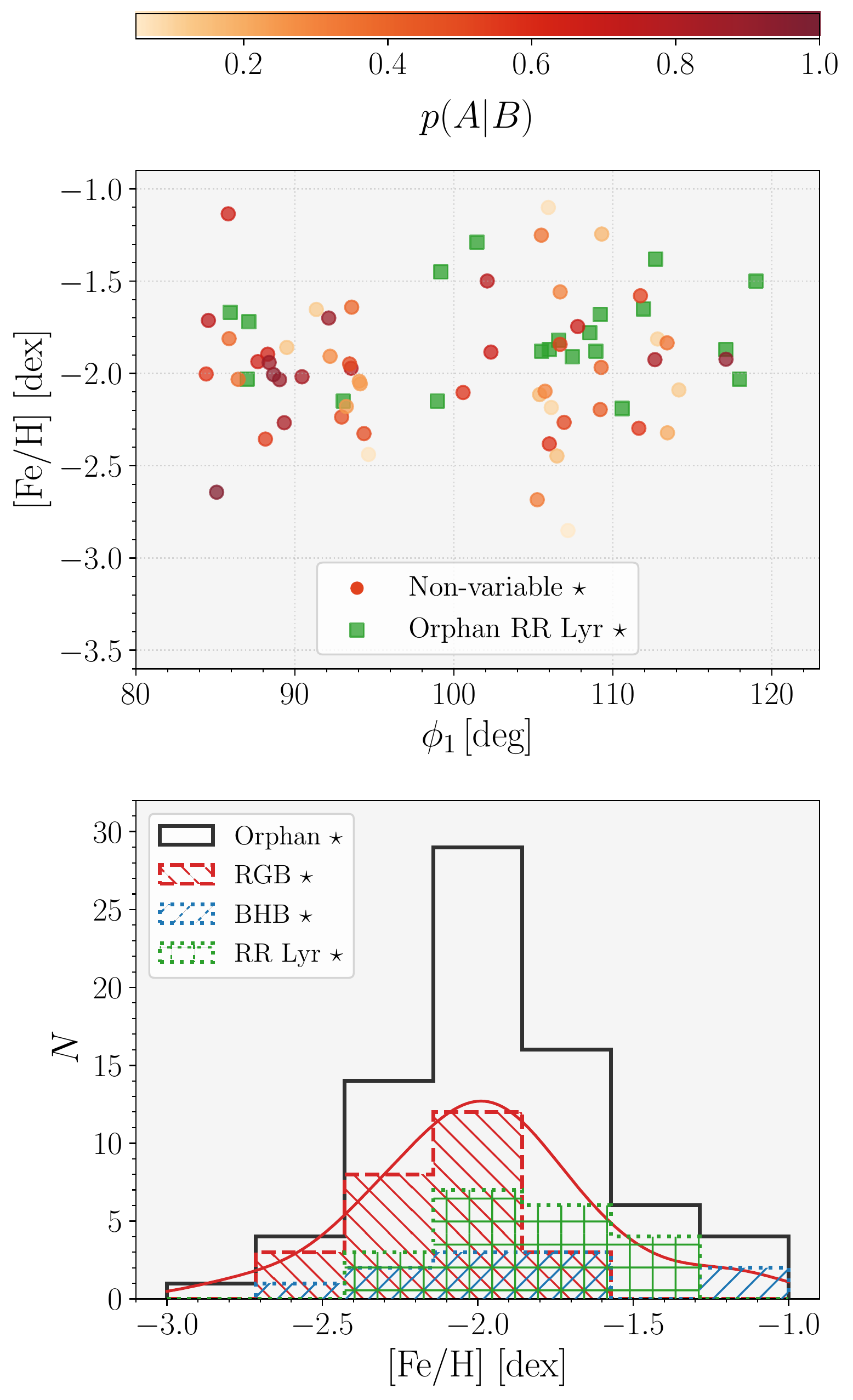}
\caption{Stream coordinates, $\phi_{\rm 1}$, versus [Fe/H] (top panel), and metallicity distribution function (bottom panel) for likely variable and nonvariable Orphan stream members. The color-coding of each nonvariable star represents the probability of association to the Orphan stream $p(A|B)$, and the green squares represent RR~Lyrae stars associated with the Orphan stream in our study. The histogram in the lower panel represents the metallicity distribution of the entire sample (black line) the RGB stars (red dashed line), BHB stars (blue dotted line), and RR~Lyrae stars (green lines). The red solid line represents the kernel density estimate of the metallicity distribution of the RGB stars.}
\label{fig:OrphanStabMetal}
\end{figure}

We explored the existence of a metallicity gradient in the Orphan stream using our nonvariable and RR~Lyrae sample\footnote{We verified, with a sample of 3000 RRL stars, that both the new high-resolution $\Delta$S scale and that of the SSPP pipeline metallicities agree within $-0.01$\,dex with a dispersion of $0.28$\,dex without any significant trend.}. The depiction of the metallicity versus $\phi_{\rm 1}$ can be found in Fig.~\ref{fig:OrphanStabMetal}. In both of our samples (nonvariable and RR~Lyrae sample) we do not detect any significant correlation between the sky position $\phi_{\rm 1}$ and metallicity. A similar outcome holds even when we include only stars with a high probability $p(A|B)>0.5$ for both of our samples. One of the possible reasons for this discrepancy lies in  the different metallicity calibrations between our study and \citet{Sesar2013Orphan}. In our case, we rely on the new calibration of the $\Delta$S method using metallicities determined from high-resolution spectra \citep{Crestani2020}, while \citet{Sesar2013Orphan} relied on the calibration of \citet{Layden1994} which is slightly offset compared to metallicities obtained from high-resolution spectra \citep[see, e.g.,][]{For2011chem,Chadid2017}. Another reason could lie in the metallicity scale, where \citet{Sesar2013Orphan} values lie on the \citet{Zinn1984} scale\footnote{It is worth mentioning that the \citet{Zinn1984} scale exhibits mild nonlinearity in comparison with the high-resolution studies of the MW globular clusters \citep[see fig.~9 in][]{Carretta2009}.}, while our metallicities are on a different metallicity scale \citep{Chadid2017,Sneden2017,Crestani2020}. This could shift the metallicities of \citet{Sesar2013Orphan} toward the metal-rich end by up to $0.2$\,dex \citep{For2011chem}. To conclude, using our dataset we were unable to confirm the existence of a metallicity gradient in the Orphan stellar stream.

\subsection{Grus\,II as a possible progenitor}

In the work by K19, the previously considered candidates for the Orphan progenitors, Segue\,1 and UMa\,II \citep{Fellhauer2007,Newberg2010}, were excluded based on their distance and proper motions. One viable candidate for the progenitor of the Orphan stream remained, Grus\,II, a UFD \citep[found in the DES by][]{Drlica2015}. Grus\,II matches with the Orphan stellar stream in the coordinates and proper motion space. Recently, line-of-sight velocities and chemical abundances became available for several stars associated with Grus\,II UFD \citep{Simon2020,Hansen2020}. The line-of-sight velocities center on average at $-106.7 \pm 0.2$\,km\,s$^{-1}$ for three RGB stars analyzed by \citet{Hansen2020}, and at $-110.7 \pm 0.5$\,km\,s$^{-1}$ for identified members by \citep{Simon2020}. Combining the distance and sky position of Grus\,II \citep{MartinezVazquez2019}, together with the proper motions \citep{McConnachie2020Dwarfs} and line-of-sight velocities \citep{Simon2020} allowed us to calculate the orbital properties of Grus\,II, and to compare them with the orbital properties of our RR~Lyrae sample associated with the Orphan stellar stream.

\subsubsection{Dynamical association}

For the purpose of examining the kinematical distribution of the identified Orphan stream members and Grus\,II, we utilized the \texttt{galpy v1.6}\footnote{Available at \url{http://github.com/jobovy/galpy}.} package for Galactic dynamics \citep{Binney2012Act,Bovy2013Staeckel,Bovy2015}, and estimated for the entire RR~Lyrae sample and Grus\,II the following quantities: orbital parameters (eccentricity $e$, excursion from the Galactic plane $z_{\text{max}}$, and peri- and apocenters, $r_{\text{per}}$ and $r_{\text{apo}}$), orbital energy $E$, actions $J_{\rm R}$, $J_{\rm z}$, and angular momenta $L_{\rm z}$ ($J_{\phi}$) with their respective uncertainties and correlations. 

In our setup, we implemented an MW potential consisting of a Miyamoto-Nagai disk \citep[$M_{\text{disk}}=6.8 \times 10^{10}$\,M$_{\odot}$, $a=3.0$\,kpc, $b=0.28$\,kpc,][]{Miyamoto1975}\footnote{For details on the disk potential see \citet{Bovy2015}.}, a Hernquist bulge \citep[][$M_{\text{bulge}} = 6.0 \times 10^{9}$\,M$_{\odot}$, $a=0.5$\,kpc]{Hernquist1990}; and a Navarro-Frenk-White spherical halo \citep[][$M_{\text{halo}}=5.4 \times 10^{11}$\,M$_{\odot}$, $r_{\text{s}}=16$\,kpc]{Navarro1997}. 

\begin{figure}
\includegraphics[width=\columnwidth]{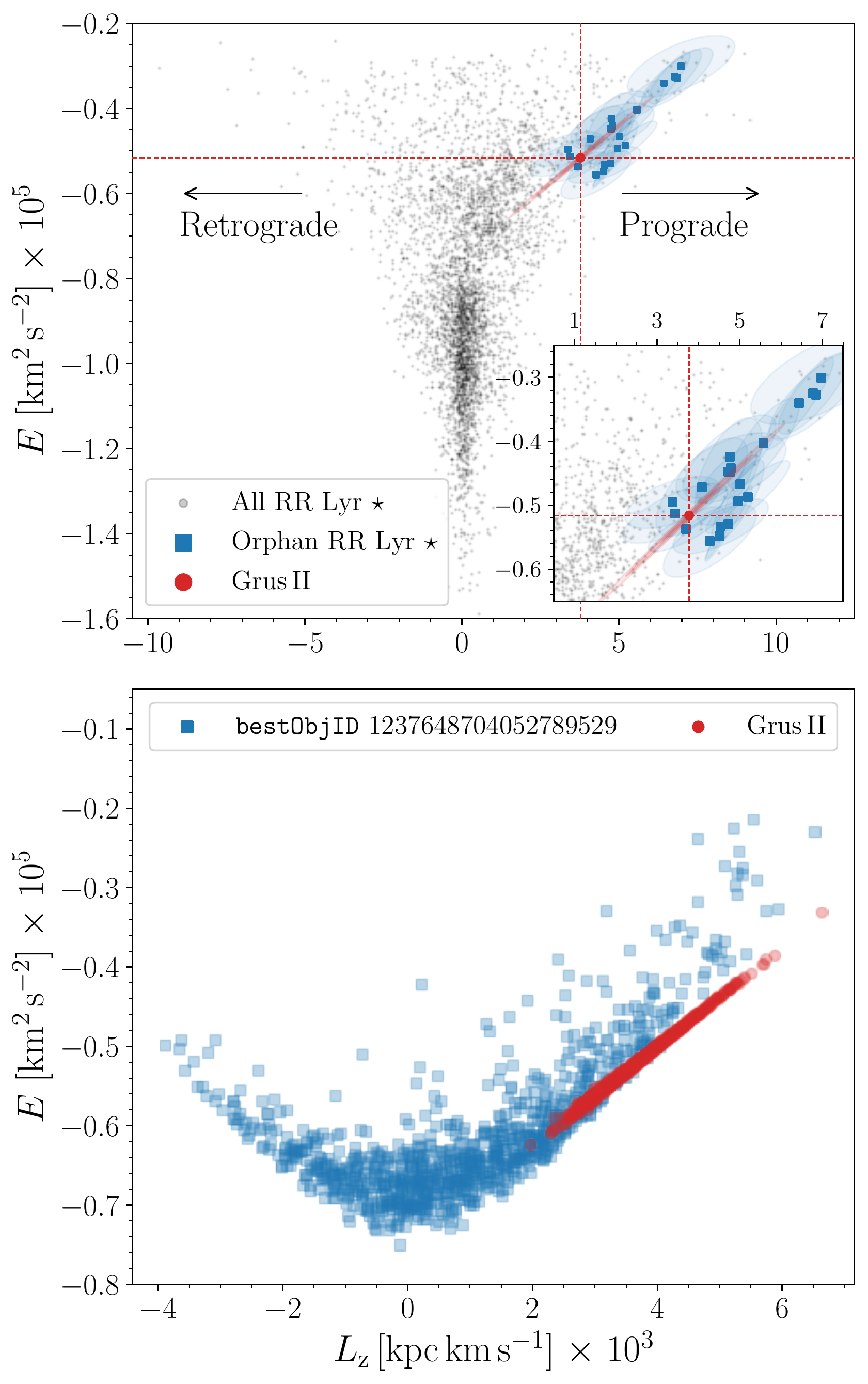}
\caption{Distribution of the orbital energy $E$ vs. the $z$-component of the angular momentum $L_{\rm z}$ (top panel). The bottom panel shows an example of the multivariate non-normal distribution of energies $E$ vs. angular momenta for one of the stars from our sample (represented with blue squares) and Grus\,II (denoted with red circles). The underlying gray points in the upper panel represent the entire RR~Lyrae sample fulfilling the condition in Eq.~\ref{eq:PMcut}. The blue squares represent the RR~Lyrae variables associated with the Orphan stellar stream. Each point is accompanied with an error ellipse estimated based on our Monte Carlo simulation. The position of Grus\,II is marked with the red dot and dashed lines accompanied by error ellipses representing the covariances.}
\label{fig:AngE}
\end{figure}

As a Galactocentric reference frame, we adopted the left-hand annotation with the following values for the Solar position and motion: The distance to the Galactic center is set to $R_{\rm 0} = 8.178$\,kpc \citep{DistantoSMBH2019}, the Solar system is placed above the Galactic plane at $z_{\odot} = 20.8$\,pc \citep{Bennett2019}. The Solar motion with respect to the local standard of rest is $\left(U_{\odot}, \boldsymbol{\upsilon_\odot},W_{\odot} \right) = \left(-11.1, 247.24, 7.25\right)$\,km\,s$^{-1}$ \citep{Schonrich2010,Schonrich2012}, where $V_{\odot} = \boldsymbol{\upsilon_\odot} - V_{\rm c} = 12.24$\,km\,s$^{-1}$. For each star we performed a Monte Carlo simulation taking into account the full covariance between the sky positions $\alpha$, $\delta$, and proper motions $\mu_{\alpha^{\ast}},~\mu_{\delta}$, in combination with errors in systemic velocities and distances. The estimated values were taken as an average of the generated distributions with the standard deviation representing the uncertainties on the given properties. In addition, to robustly assess the distributions of the orbital parameters, we also recovered the correlations between the individual orbital properties. Here we note that $E$ and actions often do not follow the multivariate normal distribution, as shown, for example, in figure~6 in \citet{Hanke2020}, and here in the bottom panel of Fig.~\ref{fig:AngE}. Thus, our assumption based on averages, standard deviations, and correlations here serves only to guide the eye and give an intuition on the uncertainties of estimated parameters. 

The median pericentric distances of RR~Lyrae stars associated with the Orphan stellar stream peak at $22$\,kpc. They reach on their orbit a median apocenter equal to $89$\,kpc, and their average eccentricity varies around $0.61$. These values are similar to the orbital properties estimated by \citet{Newberg2010}, who estimated eccentricities of Orphan stream stars to be $0.7$ with apocentric and pericentric distances equal to $\approx90$\,kpc, and $16$\,kpc, respectively. In the case of Grus\,II, the UFD reaches apocentric and pericentric distances of $66$\,kpc, and $27$\,kpc, respectively. Our calculated orbital parameters are by a construction similar to orbital properties obtained by \citet{Simon2020} since we used the same the distance and line-of-sight velocity of the Grus\,II. Its orbit has an eccentricity of $0.44$, somewhat different from the \rrlyr~RR~Lyrae stars associated with the Orphan stream in our study. In addition, looking at the best-fitting model of the Orphan orbit obtained by \citet[][see their figure~3 for reference]{Erkal2019} Grus\,II at $\phi_{\rm 1} = -66.1$\,deg, if considered as the Orphan progenitor, should have largely different line-of-sight velocity than it was measured by \citet{Hansen2020}, but further examination is highly desirable. 

Some orbital properties of likely Orphan stream members are examined in the $E$ -- $L_{\rm z}$ plane and are displayed in Fig.~\ref{fig:AngE}. All of Orphan RR~Lyrae stars clusters on positive values of $L_{\rm z}$ denoting its prograde orbit \citep[thus confirming previous findings by, e.g.,][]{Newberg2010}, and high-energy region. Grus\,II falls right in the middle of our distribution of RR~Lyrae stars, partially supporting the hypothesis of Grus\,II being the progenitor of the Orphan stellar stream. Unfortunately, large uncertainties in actions and energies prohibit a sensible comparison in the multivariate parameter space between Orphan RR~Lyrae stars and Grus\,II. At the current error budget, multivariate analysis in the action space would lead to a large number of false-positive candidates for membership with Grus\,II. 

\subsubsection{An elusive chemical connection between the Orphan stream and Grus\,II}

The broad [Fe/H] distribution of the Orphan stream supports its likely origin from a dwarf-like galaxy as was pointed out by several previous studies \citep[e.g.,][]{Sesar2013Orphan,Casey2013,Casey2014,Koposov2019Orphan,Fardal2019}. The work by \citet{Casey2013,Casey2014} used low- and high-resolution spectra of K-giants to study the chemical and kinematical properties of the Orphan stream. Using \gaia~proper motions, we were able to clean the K-giants sample from the obvious outliers using our method described in Sect.~\ref{sec:Method} and the K19 RR~Lyrae sample as a reference. We note that we did not use the radial velocities determined in our study, since they do not cover the coordinate region examined by \citet{Casey2013,Casey2014}, thus our membership probabilities are only based on coordinates and proper motions.

We found that from both studies \citep{Casey2013,Casey2014} only two stars\footnote{Marked as OSS-7 and OSS-8 in \citet{Casey2013,Casey2014}.} can be considered as likely members (having set the $p(A|B) > 0.05$ threshold). Similarly to the proper motion membership provided by \citet{Fardal2019}, we associate star OSS-8 with the Orphan stream. Unlike \citet{Fardal2019} we do not associate OSS-6 and OSS-14 with the stream given their low $p(A|B) = 0.02$ and $p(A|B) = 0.0$, respectively. This does neither significantly affect the observed metallicity spread, nor the assumed peak in its distribution. The two stars associated with the Orphan stream in our analysis exhibit very different metallicities namely; $-2.82, -1.62$\,dex \citep[based on tab.~1 in][]{Casey2013} covering the entire metallicity domain described in \citet{Sesar2013Orphan} or covered by our sample of nonvariable stars (see Sect.~\ref{sec:stableOrphan} and Fig.~\ref{fig:OrphanStabMetal} for details). 

The spectroscopic study by \citet{Hansen2020} provides a detailed abundance analysis for three likely Grus\,II members located on the RGB. The low number of stars with extensive abundance patterns associated both with the Orphan stellar stream and the Grus\,II dwarf galaxy prohibits any detailed chemical tagging. Nevertheless, the iron abundance $\text{[Fe/H]} = (-2.49; -2.69; -2.94)$\,dex for three red giants linked with Grus\,II permits a tentative discussion about their possible connection with the Orphan stream on the basis of its metallicity distribution. The metallicities of the three Grus\,II giants fall onto the metal-poor end of Orphan's metallicity distribution as traced by several independent sources: K-giants, RR~Lyrae stars \citep{Casey2013,Sesar2013Orphan}, and our RR~Lyrae and nonvariable stellar sample.

In general, the UFDs are almost exclusively old and metal-poor. On the other hand, considering a rather massive dwarf galaxy, it is expected to undergo a few episodes of star formation. This will result in stars with higher metallicities being centrally concentrated (due to past and/or ongoing star formation), while the more metal-poor stars are distributed all over the galaxy \citep{Harbeck2001,Grebel2003,Crnojevic2010,Lianou2010,Hendricks2014}. Thus, when a given dwarf enters a parent galaxy potential, it is subdued by the strong gravitational forces, which inevitably results in a tidal disruption of its peripherals, and later the dwarf itself. The outlined paradigm leads to the formation of a metallicity gradient, where metal-poor stars are stripped first followed by the metal-rich core. Such a metallicity gradient has been reported, for example, in the Sagittarius dwarf and stream \citep[see, e.g.,][]{Bellazzini1999DwarGrad,McDonald2013DwarGrad,Hayes2020SagStrGrad}. We note that dwarf galaxies with inverse metallicity gradients have been observed in other galaxy systems and at higher redshifts \citep[e.g.,][]{Wang2019,Grossi2020} but so far not in the Local Group. 

Concerning the presumed metallicity distribution between the stream and its progenitor, we assessed the probability of observing three metal-poor red giants with respect to the metallicity distribution of the Orphan stream. We employed the Gaussian kernel density estimate (KDE) from the \texttt{scikit-learn} library \citep{Pedregosa2012} to describe the aforementioned metallicity distribution. Using the \texttt{GridSearchCV} module from the \texttt{scikit-learn} library, with 10-fold cross-validation, we selected the most suitable bandwidth ($0.176$) of the Gaussian kernel for the metallicity distribution of the RGB stars associated with the Orphan stream. The resulting KDE is displayed in Fig.~\ref{fig:OrphanStabMetal}. Using the estimated metallicity KDE, we randomly drew three values simulating the random pick in observing three red giants in Grus\,II. We searched for instances where we would pick three stars with [Fe/H]$ < -2.4$\,dex. Based on one million evaluations, such an event happened only in approximately $0.2$ percent of the cases. We note that a similar results holds even when we assume $\text{[Fe/H]} = -2.51 \pm 0.11$\,dex from \citet{Simon2020} based on metallicities estimated from the Calcium triplet. Thus, connecting the Grus\,II with the Orphan stream is rather unlikely. Taking into consideration the discrepancy between metallicities in high-resolution studies and SDSS stellar parameters \citep{Smolinski2011sspp}, we would expect this probability to go even lower.

\section{Summary} \label{sec:Summary}

In this study, we presented our sample of \samp~halo RR~Lyrae stars with an available 7D chemo-dynamical distribution based on the SDSS survey, mapping mainly the northern hemisphere from four out to $100$\,kpc. We employed our dataset to study the Orphan stellar stream with which we found \rrlyr~single mode RR~Lyrae stars spatially and kinematically associated. We provide the full spatial and kinematical distribution for the identified stream members together with their spectroscopic metallicities. The average metallicity of our Orphan RR~Lyrae members centers at $-1.80(6)$\,dex, thus yielding a higher metallicity than previously reported for RR~Lyrae variables linked to the Orphan stream \citep[e.g.,][]{Sesar2013Orphan}. A higher average metallicity and the extended metallicity distribution could potentially shift the predicted mass of the Orphan progenitor from $10^6$ to $10^7$\,M$_{\odot}$ \citep[using the mass-metallicity relation from][]{Kirby2013}. Unfortunately, large uncertainties in systemic velocities of our RR~Lyrae sample prevent us from exploring the progenitor mass for the Orphan stellar stream by means of its velocity dispersion.

Using the newly identified stream members and their line-of-sight velocities, we searched for additional nonvariable members using the spectral catalog of the SDSS survey processed by the SSPP pipeline. We found additional \stab~nonvariable stars that are mainly RGB and BHB stars exhibiting different metallicity distributions $-2.13 \pm 0.05$\,dex, and $-1.87 \pm 0.14$\,dex, with dispersions of $0.23$ and $0.43$\,dex, respectively. 

The 7D chemo-dynamical distribution of the associated RR~Lyrae and nonvariable stars permitted us to carry out comparison between likely Orphan stream members and a possible Orphan progenitor, Grus\,II, a UFD discovered in the DES \citep{Drlica2015,DES2018DR1}. Kinematically, RR~Lyrae members and Grus\,II match in action and energy space, albeit with large uncertainties in the aforementioned parameters. The orbital properties also fit, both Orphan stream stars and Grus\,II follow a prograde orbit with mildly different eccentricities ($0.4 - 0.7$), and similar pericentric and apocentric passages. Since in the interaction model between the MW, the Large Magellanic Cloud, and the Orphan stream \citep{Erkal2019}, the line-of-sight velocity of Grus\,II does not exactly match, further investigation is called for. From the chemical perspective, using [Fe/H] from a study of three RGB stars by \citet{Hansen2020}, Grus\,II presumably lies on the metal-poor end of the metallicity distribution of the Orphan stream. Furthermore, considering Grus\,II as the progenitor of the Orphan stream would result in an inverse metallicity gradient between the stream and Grus\,II which would be unexpected, although we note that such dwarf galaxies \citep[stellar masses below $10^{9.5}$\,M$_{\odot}$,][]{Grossi2020} have been observed outside the Local Group. Dwarf galaxies with an inverted metallicity gradient have been found in, for example, the Virgo cluster \citep{Grossi2020} or at high redshifts \citep{Cresci2010,Queyrel2012,Wang2019}. Thus, linking Grus\,II with the Orphan stream on metallicity alone is dubious. For the reasons above we conclude that the link between Grus\,II and the northern part of the Orphan stream is rather unlikely.

This conclusion leaves us with two possible options to contemplate about the Orphan's stream's progenitor. One suggests that it has been already dissolved during its passage through the MW halo while the second option points toward the progenitor currently being located in the Galactic plane where high extinction severely hampers the efforts in search for MW satellites. Using our Gaussian process regressor between the equatorial coordinate $\alpha$ and the heliocentric distance, we looked at the expected orbit of the Orphan stream. It passes behind the Galactic plane around $d=(18 \pm 3)$\,kpc which places the stream right on the edge of the assumed MW disk. Although the currently assumed mass of the Orphan progenitor (from $10^6$ to $10^7$\,M$_{\odot}$) is not enought to warp the MW disk \citep{Burke1957,Westerhout1957}, the model passes through the strong negative vertical displacement traced by the Classical Cepheids in the outer disk \citep[see the left-hand panel of figure~7 in][]{Skowron2019}.

\begin{acknowledgements}
Z.P., A.J.K.H., B.L., E.K.G., and H.L. acknowledge support by the Deutsche Forschungsgemeinschaft (DFG, German Research Foundation) -- Project-ID 138713538 -- SFB 881 (``The Milky Way System", subprojects A03, A05, A11). VFB, MF, GA, SM and PMM acknowledge the financial support of the Istituto Nazionale di Astrofisica (INAF), Osservatorio Astronomico di Roma and Agenzia Spaziale Italiana (ASI) under contract to INAF: ASI 2014-049- R.0 dedicated to SSDC. G.F. has been supported by the Futuro in Ricerca 2013 (grant RBFR13J716). Funding for the Sloan Digital Sky Survey IV has been provided by the Alfred P. Sloan Foundation, the U.S. Department of Energy Office of Science, and the Participating Institutions. MM and JPM are supported by the U.S. National Science Foundation under Grant No. AST-1714534. We thank Marina Rejkuba for the useful discussion and thoughtful comments that helped improve the manuscript. SDSS-IV acknowledges support and resources from the Center for High Performance Computing  at the University of Utah. The SDSS website is www.sdss.org. SDSS-IV is managed by the Astrophysical Research Consortium for the Participating Institutions of the SDSS Collaboration including the Brazilian Participation Group, the Carnegie Institution for Science, Carnegie Mellon University, Center for Astrophysics | Harvard \& Smithsonian, the Chilean Participation Group, the French Participation Group, Instituto de Astrof\'isica de Canarias, The Johns Hopkins University, Kavli Institute for the Physics and Mathematics of the Universe (IPMU) / University of Tokyo, the Korean Participation Group, Lawrence Berkeley National Laboratory, Leibniz Institut f\"ur Astrophysik Potsdam (AIP),  Max-Planck-Institut f\"ur Astronomie (MPIA Heidelberg), Max-Planck-Institut f\"ur Astrophysik (MPA Garching), Max-Planck-Institut f\"ur Extraterrestrische Physik (MPE), National Astronomical Observatories of China, New Mexico State University, New York University, University of Notre Dame, Observat\'ario Nacional / MCTI, The Ohio State University, Pennsylvania State University, Shanghai Astronomical Observatory, United Kingdom Participation Group, Universidad Nacional Aut\'onoma de M\'exico, University of Arizona, University of Colorado Boulder, University of Oxford, University of Portsmouth, University of Utah, University of Virginia, University of Washington, University of Wisconsin, Vanderbilt University, and Yale University. This work has made use of data from the European Space Agency (ESA) mission {\it Gaia} (\url{https://www.cosmos.esa.int/gaia}), processed by the {\it Gaia} Data Processing and Analysis Consortium (DPAC, \url{https://www.cosmos.esa.int/web/gaia/dpac/consortium}). Funding for the DPAC has been provided by national institutions, in particular the institutions participating in the {\it Gaia} Multilateral Agreement. The CSS survey is funded by the National Aeronautics and Space Administration under Grant No. NNG05GF22G issued through the Science Mission Directorate Near-Earth Objects Observations Program. The CRTS survey is supported by the U.S.~National Science Foundation under grants AST-0909182 and AST-1313422.

This research made use of the following Python packages: \texttt{Astropy} \citep{astropy2013,astropy2018}, \texttt{dustmaps} \citep{Green2018dust}, \texttt{emcee} \citep{Foreman-Mackey2013}, \texttt{galpy} \citep{Bovy2015}, \texttt{IPython} \citep{ipython}, \texttt{Matplotlib} \citep{matplotlib}, \texttt{NumPy} \citep{numpy}, \texttt{scikit-learn} \citep{Pedregosa2012}, and \texttt{SciPy} \citep{scipy}.
\end{acknowledgements}

\bibliographystyle{aa}
\bibliography{biby} 

\begin{appendix} 
\section{Processing the photometric data from CSS} \label{ap:Photometry}

\subsection{Processing known RR Lyrae in CSS and \gaia}

\begin{figure}
\includegraphics[width=\columnwidth]{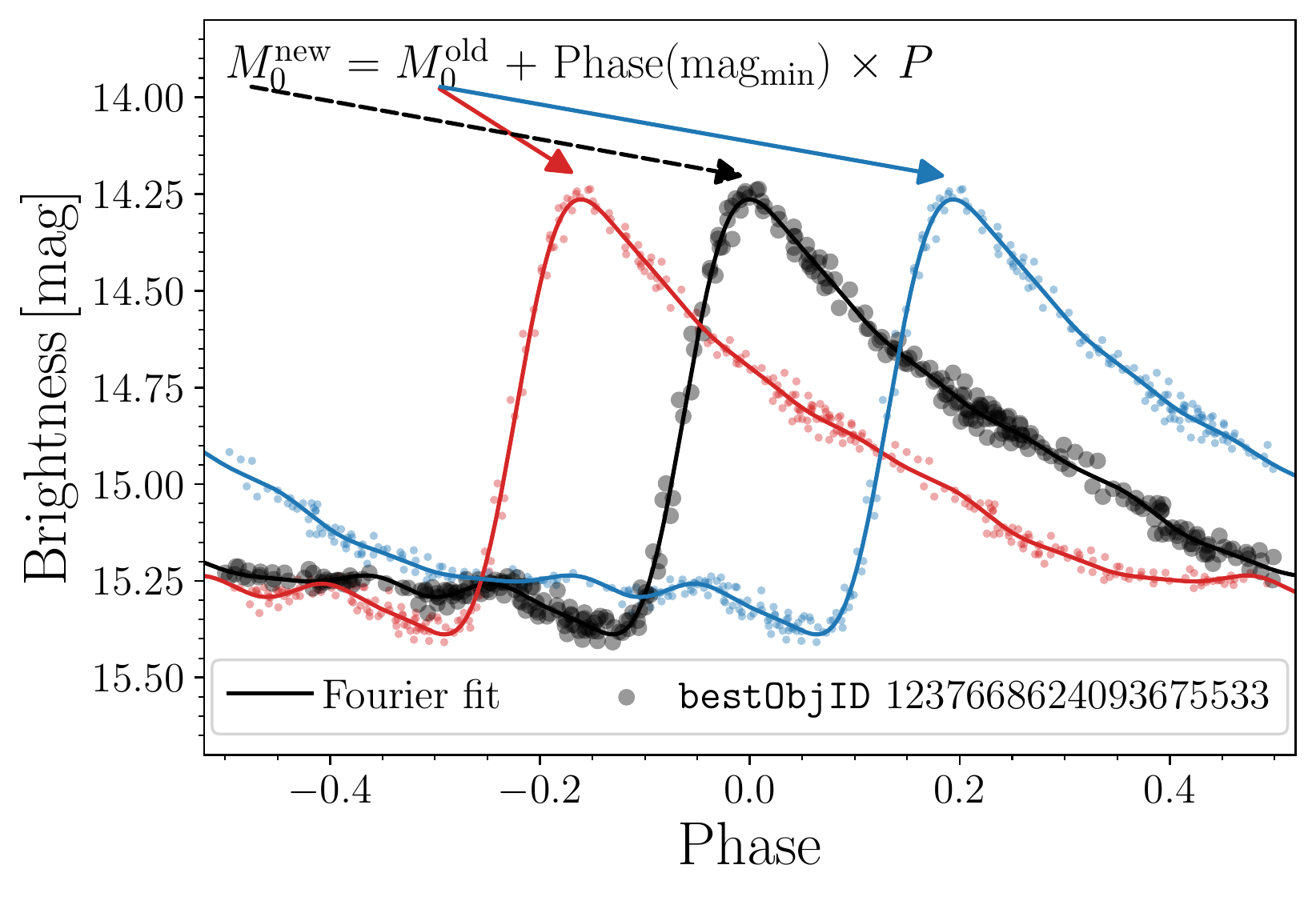}
\caption{Example of the $M_{0}$ determination based on CSS photometry for one of the sample stars. The blue and red dots represent erroneous $M_{0}$ that were subsequently corrected by a phase shift of the time of brightness maxima (determined from the Fourier fit) multiplied by the pulsation period.}
\label{fig:LCm0}
\end{figure}

Our initial step in the verification of our sample was to establish the dominant pulsation period. Thus, we retrieved the pulsation periods for stars in our sample that were identified as RR~Lyrae stars both in CSS and \gaia~EDR3 \citep{Drake2013,Drake2013stream,Drake2014CatVari,Abbas2014,Clementini2019}, and compared their pulsation periods. When the difference between periods in \gaia~and CSS was larger than 0.005\,days, we performed a period analysis using the \texttt{Period04} software \citep{Lenz2004Period04} on the CSS data in order to establish the dominant period. Once the variability periods $P$ were secured, we focused on the determination of the time of brightness maxima $M_{0}$. We proceeded iteratively: first, we phased the retrieved CSS light curves using the determined periods and as a time of brightness maxima we selected the brightest point on the light curve. In the second step we decomposed the light curves using the Fourier decomposition:
\begin{equation} \label{eq:FourSer}
m\left ( t \right ) = A_{0} + \sum_{i=1}^{n} A_{i} \cdot \text{cos} \left (2 \pi k \left(\text{MJD}-M_{0}\right)/P + \varphi_{i} \right ) \\,
\end{equation}
where $\varphi_{i}$ and $A_{i}$ stand for phases and amplitudes, and MJD represents the Modified Julian Date at the time of observation, and $A_{0}$ represents the mean magnitude. The optimal degree, $n$, of the Fourier decomposition was estimated by gradually increasing the order until the condition on Fourier amplitude was broken $A_{i}/\sigma_{i} > 4$. From the Fourier fit, we determined the phase of the brightest point and added its period-corrected value from the initial $M_{0}^{\text{init}}$ creating a new, updated $M_{0}^{\text{upd}}$ which entered again in the first step (see an example in Fig.~\ref{fig:LCm0}). After a few iterations (usually up to 5) we derived a final time of brightness maxima. We note here that for the subsequent spectroscopic analysis (see Sect.~\ref{ap:SpectraRad}) we favored $M_{0}$ determined from the analysis of CSS data due to a larger number of observations (as compared to \gaia), and because the CSS photometric observations were conducted roughly at the same time as the SDSS observations. This ensured a consistent classification of our sample since RR~Lyrae stars can rapidly change their pulsation mode within a few years \citep[see; e.g.,][]{Soszynski2017BulgeCep}. Furthemore, strong period changes \citep[especialy in the first-overtone pulsators, see, i.e.,][]{Jurcsik2001,Szeidl2011} can introduce an additional source of uncertainty in the determination of $M_{0}$.

In the next step, we visually verified the variability of the individual phased light curves using the CSS photometry, and we removed stars with no signs of luminosity variation. Alongside this step, using a Fourier decomposition, we determined basic light curve parameters for the RR~Lyrae sample, for example, pulsation amplitudes $\text{Amp}^{V_{\rm CSS}}$\footnote{Defined as a magnitude difference between the faintest and brightest point of the Fourier fit.}, rise time $\text{RT}^{V_{\rm CSS}}$\footnote{Determined from the Fourier fit as a phase difference between the brightest and faintest point.}, amplitude ratios ($R_{\rm 21},~R_{\rm 31}$) and phase differences ($\varphi_{\rm 21},~\varphi_{\rm 31}$) defined as follows:
\begin{equation} \label{eq:FourCoe}
\varphi_{i1} = \varphi_{i} - i\varphi_{1}  \\
R_{i1} = \frac{A_{i}}{A_{1}} \\.
\end{equation}
The estimated photometric parameters allowed us to robustly assess the pulsation subclasses (RRab, RRc, and RRd) of the studied RR~Lyrae stars. Based on their position in the period-amplitude diagram and amplitude ratio vs. pulsation period, we divided them into the two categories RRab and RRc\footnote{We note that we identified some RRd pulsators but they were not used in our study.}. Variables on the borderline between both classes were examined further using an automated routine that removed the dominant pulsation mode and searched for signs of an additional mode that would coincide with a period ratio typical for double-mode RR~Lyrae stars \citep[$P_{\rm 1O} / P_{\rm F}$ from $0.68$ to $0.76$,][]{Smolec2015-BL,Soszynski2016aRRd,Prudil2017a}. In the end, variables with signs of double-mode behavior were removed from our sample.

\subsection{Searching for new RR~Lyrae stars in CSS data}

Taking advantage of the extensive SDSS sample and available CSS photometry, we conducted a new search for RR~Lyrae stars, similar to the one performed in \citet{Hanke2020}. As an initial step, we removed stars that did not have an effective temperature determined using the SSPP pipeline, assuming that they are extragalactic sources. In a second step, we looked at the color space of our confirmed RR~Lyrae sample, using SDSS multi-band photometry. Based on their color distribution, we applied rectangular color cuts on the entire SDSS spectral sample:
\begin{gather}
-1.0 < (u - g) < 1.4 \\
-0.1 < (g - r) < 0.35 \\
-0.1 < (r - i) < 0.15 \\
-0.15 < (i - z) < 0.15 \hspace{1cm}.
\end{gather} 
We note that our color conditions are similar to the ones used by \citet{Sesar2010} and \citet{Abbas2014}, only more restricted. In addition, we did not use dereddened magnitudes. 

Using the sample selected on the SDSS products, we retrieved their CSS photometry and searched for signs of variability using the \texttt{upsilon} software package\footnote{Accessible at: \url{https://github.com/dwkim78/upsilon}.} \citep{Dae-Won2016}. This software searches for variability in the provided photometric data and yields a classification (and class probability) of the variable objects based on the shape of their light curves. To ensure a correct classification, we selected for further examination only stars marked as RR~Lyrae stars with a class probability above 50\,\%. Then, using the determined pulsation periods from the \texttt{upsilon} package we determined $M_{0}$ (as described above), visually verified their periodicity in the phased light curves and removed the misclassified stars. For the final (pure) sample, we determined the Fourier coefficients and classified RR~Lyrae in subclasses.

\begin{figure}
\includegraphics[width=\columnwidth]{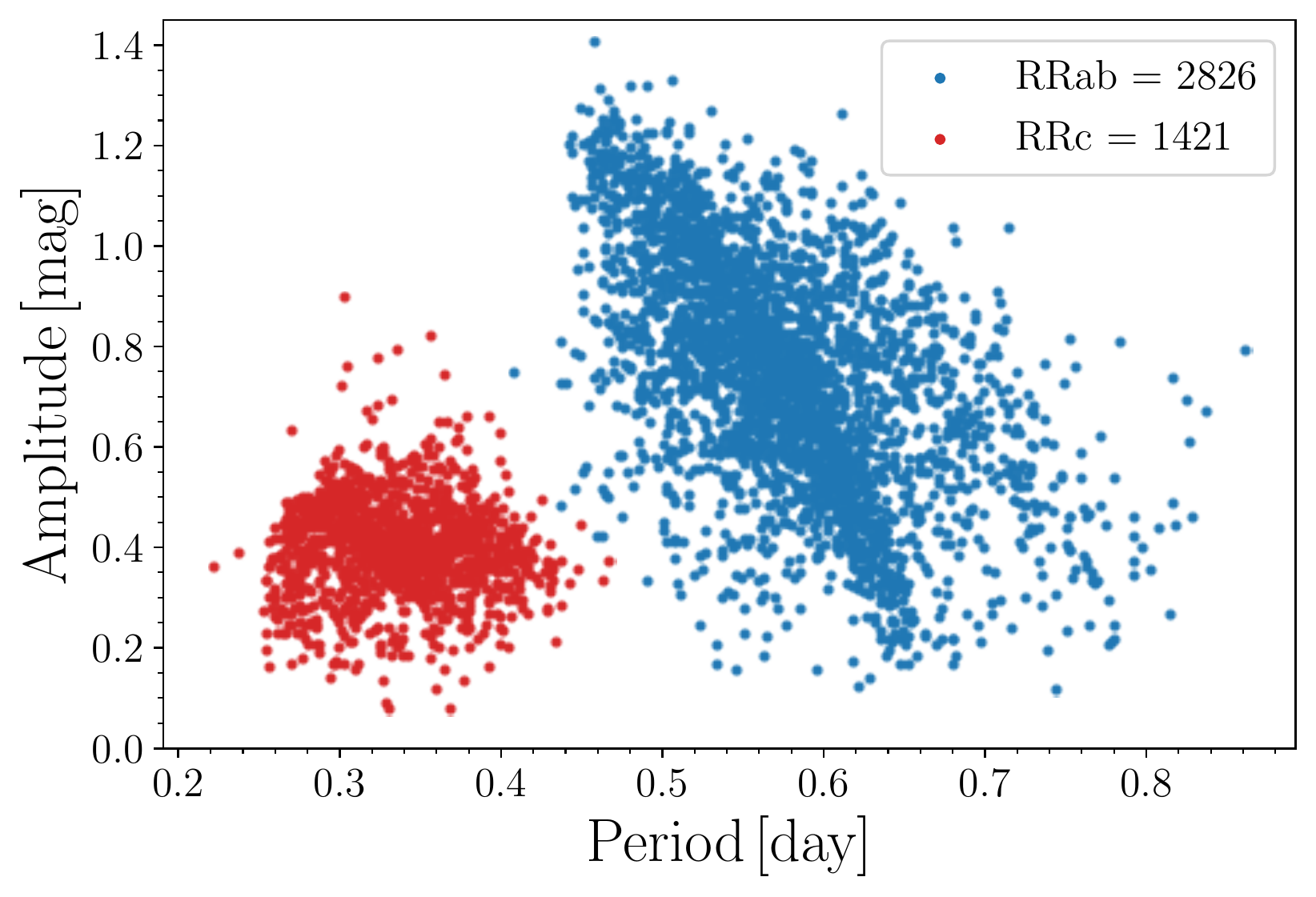}
\caption{Period-amplitude diagram for the studied sample of RR~Lyrae stars. Blue and red dots represent the fundamental and first overtone pulsators, respectively.}
\label{fig:PAdiag}
\end{figure}

As a last step, we cross-matched our sample of RR~Lyrae stars with the PanSTARRS-1 (PS1) survey catalog of RR~Lyrae stars \citep{Sesar2017a}, where their mean magnitudes were later used for the distance estimation (see Sect.~\ref{sub:DistEst}). In the end, our total sample consists of \samp~RR~Lyrae stars (2826~RRab and 1421~RRc) with photometric, astrometric, and spectroscopic data that entered our analysis. In Fig.~\ref{fig:PAdiag} we depict the distribution of the final sample in the period-amplitude diagram.

\section{Processing the spectroscopic data from SDSS} \label{ap:SpectraRad}

Obtaining a precise systemic velocity $v_{\text{sys}}$ for a given RR~Lyrae variable is hampered by the entanglement of the measured line-of-sight velocity, $v_{\text{los}}$, and the motion of the atmosphere due to pulsation. The amplitude variation of the line-of-sight velocity curves depends on the atmosphere depth. Therefore, lines formed in the upper levels of the atmosphere (e.g., the Balmer lines H$\alpha$, H$\beta$, etc.) yield larger amplitude variation, in contrast to metallic lines from elements like Fe or Sr, which are formed lower and thereby expose smaller variations in line-of-sight velocities. The line-of-sight velocity curves measured from lines in the upper and lower layers of the atmosphere vary not only in amplitude but also in shape \citep[see, e.g., figure~1 in][]{Sesar2012}. Thus, to estimate precisely the systemic velocity of a given RR~Lyrae star one needs to follow the entire pulsation cycle or utilize line-of-sight velocity templates defined for individual spectral lines \citep[metallic lines, H$\alpha$, H$\beta$, H$\gamma$, H$\delta$, see ][for instance]{Liu1991,Sesar2012,Braga2020hopefully}. The aforementioned templates scale with the photometric amplitudes, hence one can determine the systemic velocity using a single spectral line, the time of the observation, ephemerides, and amplitude information from photometry.

\begin{figure}
\includegraphics[width=\columnwidth]{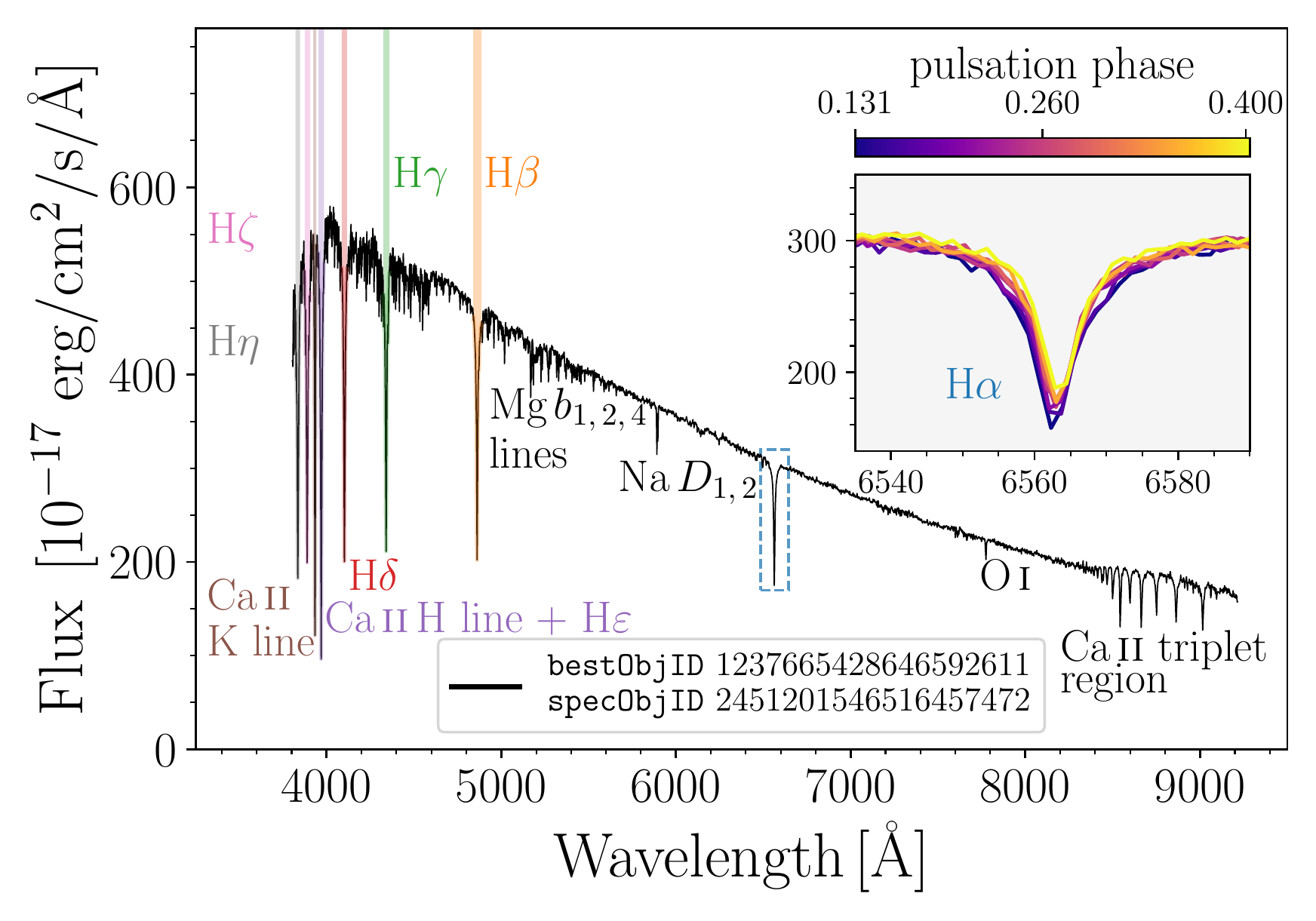}
\caption{Example of an SDSS co-added spectrum (black line) for an RR~Lyrae variable from our sample with the most prominent lines annotated. The individual exposures around the H$\alpha$ line are depicted in the inset and color-coded based on the pulsation phase.}
\label{fig:SpectExample}
\end{figure}

The available spectra from the SDSS are of low resolution ($\approx 2000$) with only a few prominent lines, mainly of the Balmer series (see Fig.~\ref{fig:SpectExample} for an example of one of our spectra) that remain detectable throughout our sample. The spectra for individual stars were obtained from the SDSS Science Archive Server\footnote{\url{https://dr15.sdss.org/sas/dr15/}}, and consist of the co-added spectra and individual exposures in both SDSS spectral windows (blue and red). Each exposure contains a header with information about the time of the observation and data composed of vacuum wavelengths\footnote{We note that for the determination of line-of-sight velocities we converted SDSS vacuum wavelengths to the air wavelength frame using a formula from \citet{Ciddor1996}.} in the heliocentric frame, flux-calibrated spectra (in units of $10^{-17}$\,erg\,s$^{-1}$\,cm$^{-2}$\,\AA$^{-1}$), and their associated errors \citep{Stoughton2002}.

To consistently estimate the systemic velocities of our RR~Lyrae sample, we proceeded in the following way. We separated the individual exposures (blue and red part of the spectrum) and selected four prominent Balmer lines (H$\alpha$, H$\beta$, H$\gamma$, H$\delta$) for which we determined their line-of-sight velocities by cross-correlation with a synthetic spectrum using the \texttt{iSpec} package \citep{Blanco2014,Blanco2019iSpec}. The synthesized spectra for each line were obtained through a python wrapper of the radiative transfer code MOOG \citep[February 2017 version,][]{Sneden1973}, using the ATLAS9 model atmospheres \citep{Castelli2003}, a solar reference scale from \citet{Asplund2009}, and a line list from VALD\footnote{\url{http://vald.astro.uu.se/}}, all of which are implemented in \texttt{iSpec}. The synthesized spectra were calculated with respect to a set of typical stellar parameters of RR~Lyrae stars \citep{For2011chem,Sneden2017,Preston2019}:
\begin{itemize}
\item $T_{\rm eff} = 6600$\,K
\item log\,$g = 2.25$\,dex
\item{[Fe/H]$ = -1.5$\,dex}
\item Microturbulence velocity $\xi_{\rm turb} = 3.5$\,km\,s$^{-1}$ \hspace{1cm} .
\end{itemize} 
A region ($\pm$100\,\AA) around each Balmer line was cross-correlated with the synthetic spectrum. To account for the uncertainties in the flux we employed a Monte-Carlo simulation by varying the flux within its errors (assuming that they follow a Gaussian distribution). This allowed us to identify problematic spectra and to assign their $v_{\text{los}}$ larger uncertainties than they would have using a single cross-correlation procedure. 

Using this approach, we discarded line-of-sight velocities that failed at least one of the following conditions:
\begin{equation} \label{eq:conditions}
\left | v_{\text{los}} / \sigma_{v_{\text{los}}} \right | > 2 \\ \cup \\ 
\sigma_{v_{\text{los}}} < 10~\text{km\,s}^{-1} \\.
\end{equation}
To determine the systemic velocities of our RR~Lyrae sample, we used a new set of line-of-sight velocity templates for the Balmer lines from \citet{Braga2020hopefully}, and scaled them by the provided linear scaling relations between the line-of-sight velocity amplitudes and the light curve amplitudes \citep[see][for details]{Braga2020hopefully}.

The systemic velocity for each Balmer line was estimated by minimizing the offset between the amplitude-scaled line-of-sight velocity templates and the measured line-of-sight velocities. For this process, we utilized the Markov Chain Monte Carlo (MCMC) sampler implemented in the \texttt{emcee} package \citep[\texttt{v.3.0.2}\footnote{\url{https://github.com/dfm/emcee/}.},][]{Foreman-Mackey2013} where we maximalized the posterior probability defined in the following way:
\begin{equation} \label{eq:PosterLikeliVsys}
p(\boldsymbol{\theta} | D) \propto p(\boldsymbol{\theta}) \times \prod^{N} p(D_{n} | \boldsymbol{\theta}_{n})\\,
\end{equation}
where $D_{n}$ represents data for an individual star in the form:
\begin{equation} \label{eq:DataMCMCM}
D_{n} = \left \{ P_{n}, M_{0,n}, v_{\text{los},n}^{\rm H \alpha}, v_{\text{los},n}^{\rm H \beta}, v_{\text{los},n}^{\rm H \gamma}, v_{\text{los},n}^{\rm H \delta} \right \}\\,
\end{equation}
and $\boldsymbol{\theta}$ the model consisting of an amplitude scaled line-of-sight velocity template for the individual Balmer line \citep[from][]{Braga2020hopefully}, each shifted by the systemic velocity. In our MCMC setup we therefore sampled the following model parameters:
\begin{equation} \label{eq:ModelMCMCM}
\boldsymbol{\theta}_{n} = \left \{ \Delta M_{0,n}, v_{\text{sys},n}^{\rm H \alpha}, v_{\text{sys},n}^{\rm H \beta}, v_{\text{sys},n}^{\rm H \gamma}, v_{\text{sys},n}^{\rm H \delta} \right \} \\,
\end{equation}
with $\Delta M_{0,n}$ representing the shift in the time of maximum light. This offset has been included since the photometric quality degrades at the faint end of our sample and the estimation of $M_{0}$ becomes challenging. This is particularly true for the first-overtone pulsators, where symmetrical light curves with lower amplitudes and larger photometric errors hamper the precise determination of $M_{0}$. The uncertainty of $M_{0}$ can affect the systemic velocity determination for stars with observations around the time of the brightness maxima, where the line-of-sight velocities change rapidly. Thus, the offset parameter, $\Delta M_{0,n}$, can compensate for such an eventuality. As a prior for our model parameters, we adopted uniform ($\mathcal{U}$) priors:
\begin{gather}  \label{eq:VsysPriors}
p(\boldsymbol{\theta}_{n}) = \mathcal{U}(-0.1 < \Delta M_{0,n} < 0.1) \hspace{0.5cm} \cap \\
 \hspace{1.15cm} \mathcal{U}(\bar{v}_{\text{los}}^{\rm H\,line} - 130, \bar{v}_{\text{los}}^{\rm H\,line} + 130)\hspace{1cm}, 
\end{gather} 
where $\bar{v}_{\text{los}}^{\rm H\,line}$ represents the median velocity for all lines, with the value $130$\,km\,s$^{-1}$ characterizing the maximal line-of-sight velocity amplitude for an RR~Lyrae star with $\text{Amp}^{V_{\rm CSS}} \approx 1.4$\,mag. $p(D_{n}, \boldsymbol{\theta}_{n})$ represents the likelihood for each line of a given star:
\begin{equation} \label{eq:LikeliHood}
p(D_{n}, \boldsymbol{\theta}) = \mathcal{N}(v_{\rm los}^{\rm H\,line} , \sigma_{v_{\rm los}^{\rm H\,line}} | v_{\rm los}^{\rm H\,model}) \\,
\end{equation}
where $v_{\rm los}^{\rm H\,model}$ represents a velocity value for a given phase of the observation $\vartheta = (\text{MJD} - M_{0} + \Delta M_{0}) / P$, from the amplitude-scaled line-of-sight velocity template shifted by $v_{\rm sys}^{\rm H\,line}$. 

\begin{figure*}
\includegraphics[width=2\columnwidth]{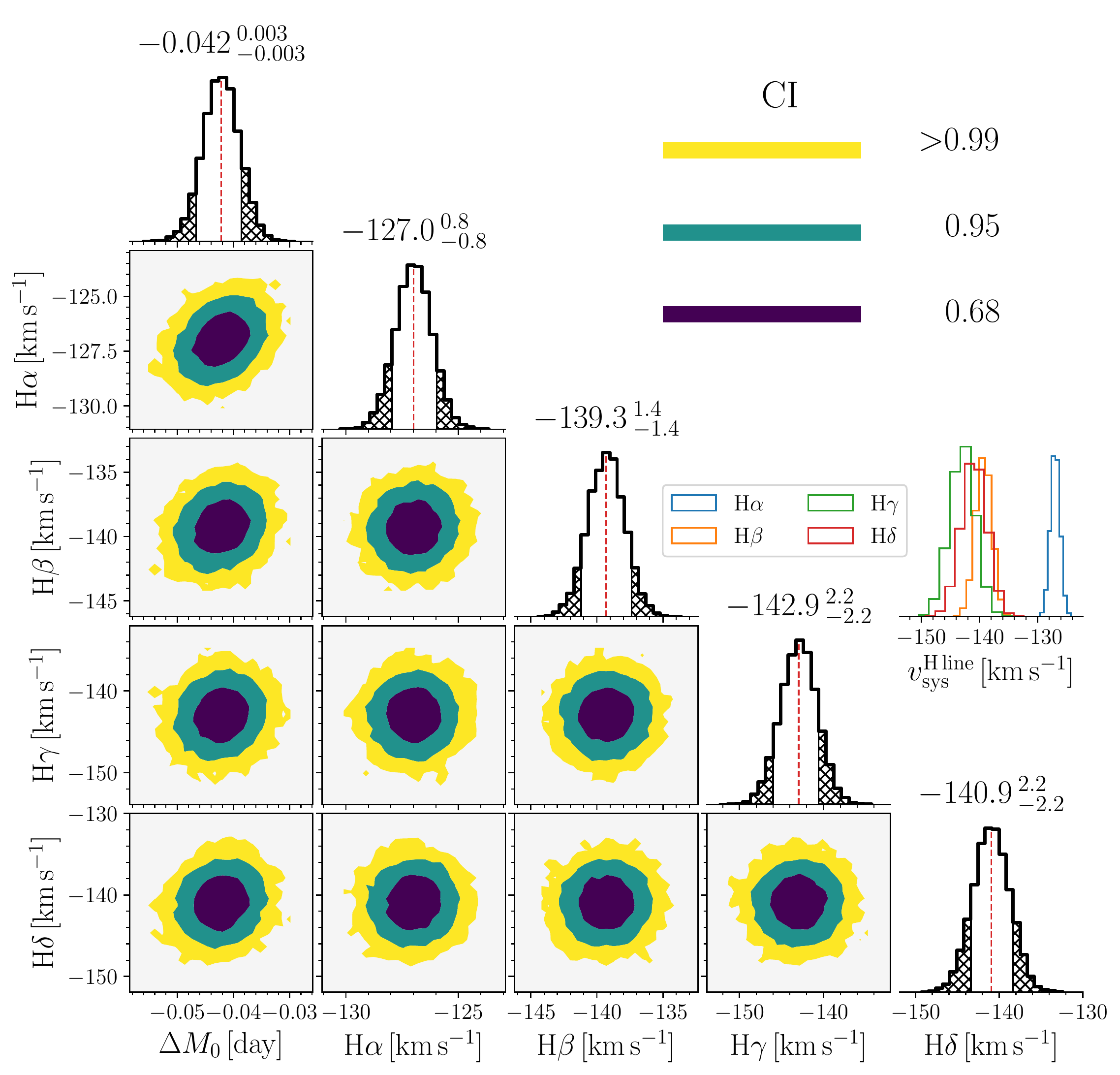}
\caption{Posterior probability distribution for parameters of our model for a given RR~Lyrae star in our sample (\texttt{bestObjID}$=1237649919501598744$) with confidence intervals (CIs) color-coded. We see an offset between systemic velocities determined from H$\beta$, H$\gamma$, and H$\delta$ in comparison with a systemic velocity determined through the H$\alpha$ line (see color histogram on the right-hand side of the figure).}
\label{fig:MCMCspec}
\end{figure*}

To estimate the posterior distribution of our model parameters, we ran \texttt{emcee} with 48 walkers for an initial 200 steps as burn-in and then restarted the sampler for an additional 2200 steps. Fig.~\ref{fig:MCMCspec} depicts the posterior likelihood distribution of the model parameters $\boldsymbol{\theta}$ for a given RR~Lyrae star from our sample. 

While examining the systemic velocities determined from individual lines, we noticed a non-negligible offset in systemic velocities between individual lines, where values determined on the blue end of the spectrum showed on average smaller values than the lines on the red end. We further examined this discrepancy in nonvariable stars\footnote{In total 162 stars covering a broad range of log\,$g \approx 3$\,dex, temperatures $\approx 4500$\,K, and metallicities $\approx 2$\,dex.} associated with three star clusters (M\,13, M\,15, and M\,67), where we performed a piecewise cross-correlation in the following way: for each exposure of a given star, we divided the spectrum into three sections based on wavelength; 
\begin{gather} \label{eq:SpecSectors}
\lambda_{1} = (4000; 4500) \AA \\
\lambda_{2} = (4500; 5000) \AA \\
\lambda_{3} = (5500; 7000) \AA \hspace{1cm}.
\end{gather}
These three wavelength regions approximately represent spectral regions covering H$\delta$ and H$\gamma$ ($\lambda_{1}$), H$\beta$ ($\lambda_{2}$), and H$\alpha$ ($\lambda_{3}$). For each part of the spectrum, we determined the line-of-sight velocity using a synthesized template spectrum generated using the SSPP pipeline-derived quantities for $T_{\rm eff}$, log\,$g$, and [Fe/H]. We found that the average line-of-sight velocities from individual exposures are decreasing as we move from the red, $\lambda_{3}$, to the blue part, $\lambda_{1}$, of the spectrum. The difference between the bluest and reddest regions is on average $-13$\,km\,s$^{-1}$. In addition, a difference between the second bluest region, $\lambda_{2}$, and the reddest, $\lambda_{3}$, region was found as well (on average $-10$\,km\,s$^{-1}$). The comparison between the known line-of-sight velocities of the three star clusters \citep[using literature values,][]{Geller2015M67,Baumgardt2019} showed that the line-of-sight velocities determined in the red region match very well literature values, while the line-of-sight velocities from the blue regions showed the aforementioned offsets. 

We decided to include this systematic offset in the determined systemic velocities for H$\gamma$, H$\delta$ (shift by $+13$\,km\,s$^{-1}$) and H$\beta$ (shift by $+10$\,km\,s$^{-1}$). The final systemic velocity value, $v_{\text{sys}}$, for a given RR~Lyrae star was estimated through a weighted average using all four Balmer lines. For its uncertainty, we adopted a weighted standard deviation $\sigma_{v_{\text{sys}}}$. On average, our weighted uncertainties are on the order of 14\,km\,s$^{-1}$. We note here that we chose to determine the systemic velocities for each line separately instead of combining them, since this approach leads to uncertainties on the systemic velocities that are considerably lower than the precision of the SDSS wavelength calibration \citep[$<5$\,km\,s$^{-1}$,][]{Lee2008SsppII, Allende2008}.

\begin{figure}
\includegraphics[width=\columnwidth]{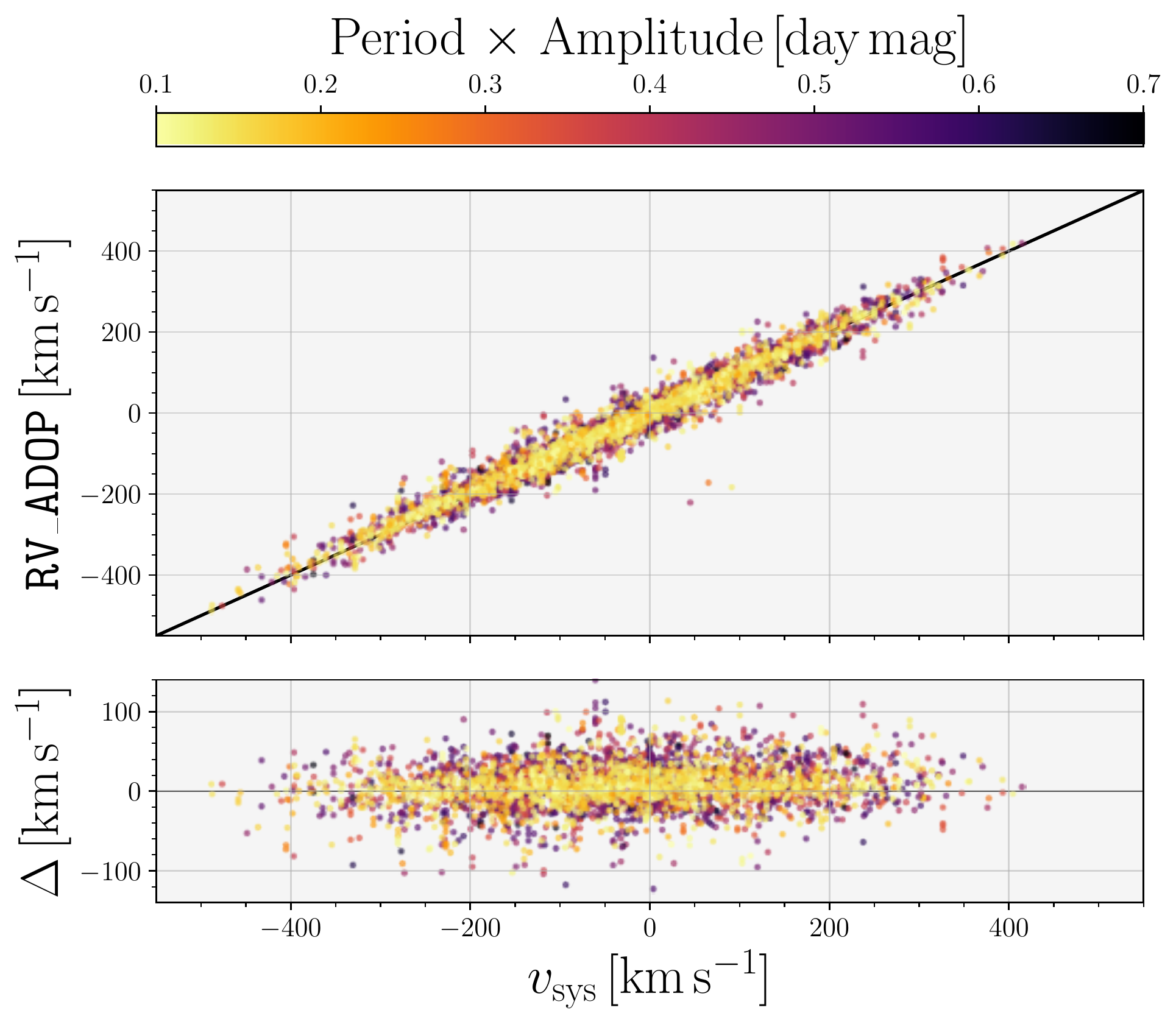}
\caption{Comparison between line-of-sight velocities \texttt{RV\_ADOP} derived by SSPP and our systemic velocities calculated using the line-of-sight velocity templates (top panel) and the residuals of their difference (bottom panel) with a color coding that is based on the product of pulsation period and amplitude.}
\label{fig:Rad-Sys-comparison}
\end{figure}

As a test for our determined systemic velocities, we compared our results ($v_{\text{sys}}$) with the heliocentric line-of-sight velocities, \texttt{RV\_ADOP}. As expected, our systemic velocities linearly follow the values from the SSPP with a substantial scatter ($\approx$29\,km\,s$^{-1}$) which is mainly caused by the pulsations of our targets and originate from erroneous estimates on the basis of coadded spectra. In Fig.~\ref{fig:Rad-Sys-comparison}, we see that stars with low amplitudes and short pulsation periods (first-overtone pulsators) exhibit a dispersion of 25\,km\,s$^{-1}$ and cluster around unity (black solid line in the top panel). In contrast, stars at the other end of the period-amplitude distribution exhibit a larger scatter since the chances of observing them around the time of mean line-of-sight velocity are lower. Fundamental pulsators exhibit a dispersion of 31\,km\,s$^{-1}$.

\section{Additional figures}

\begin{figure*}
\includegraphics[width=2\columnwidth]{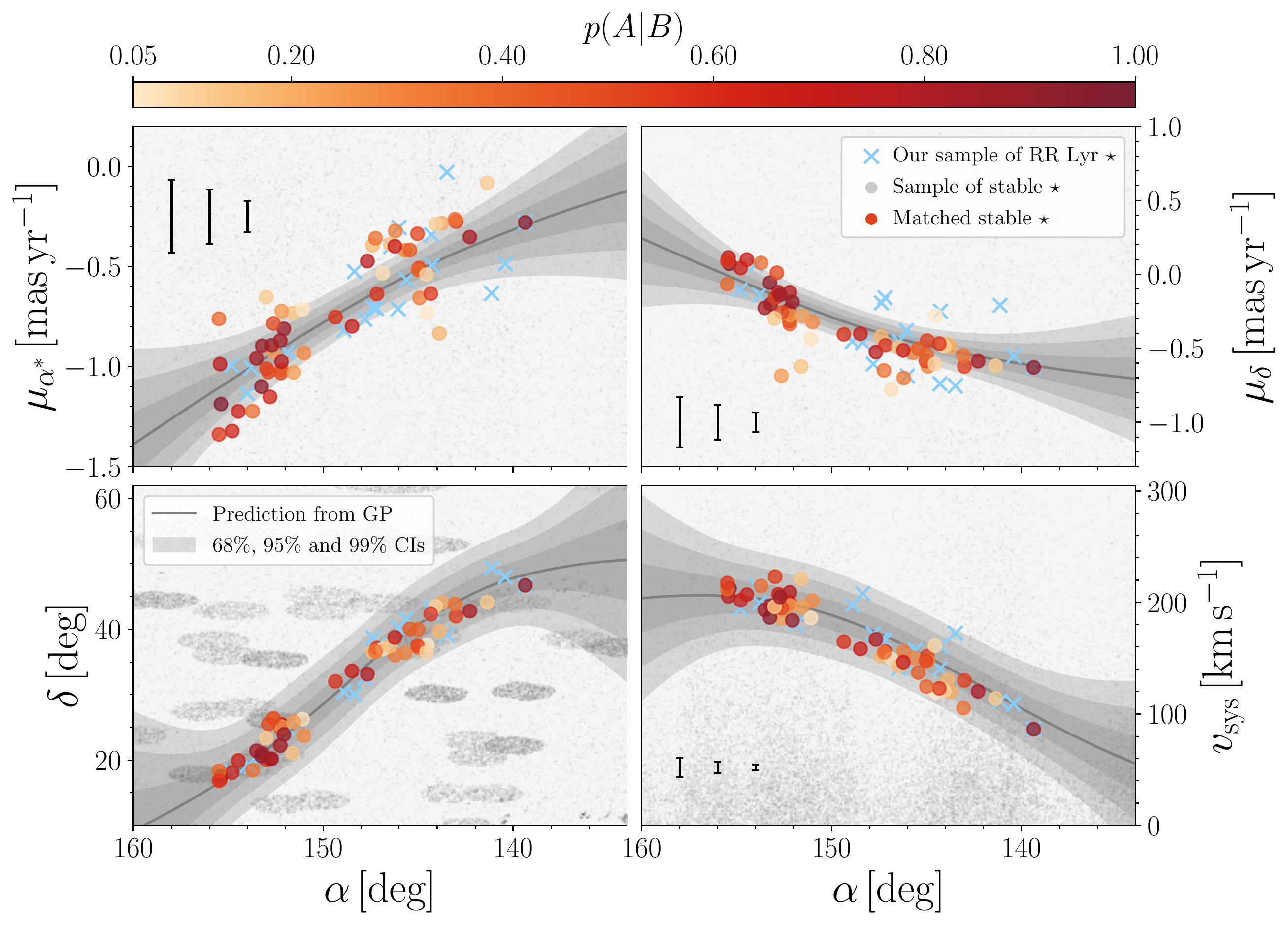}
\caption{Four-parameter association of nonvariable stars with our identified sample of RR~Lyrae variables (blue crosses) in the Orphan stellar stream. Similar to Fig.~\ref{fig:OrphanMem}, the color coding denotes the membership probabilities $p(A|B)$ in coordinate (bottom left panel), proper motion (upper panels), and systemic velocity (bottom right panel) space. The gray lines and shading represent the Gaussian process regression and confidence intervals (CIs), respectively. The three error bars at the bottom of each panel denote the $15.9$, $50$, and $84.1$ percentiles of the uncertainties on individual parameters.}
\label{fig:Orphan-stable}
\end{figure*}

\section{Additional tables}

\setlength{\tabcolsep}{1.3pt}
\renewcommand{\arraystretch}{1.2}
\begin{landscape}
\begin{table}
\caption{List of RR~Lyrae variables in our sample associated with the Orphan stream based on our analysis. The first two columns denote the SDSS and \gaia~EDR3 object IDs followed by their equatorial coordinates in columns three and four. Columns 5, 6, 7, and 8 list the estimated distances and systemic velocities with associated uncertainties. The parameters estimated on basis of the CSS photometry are listed in columns 9, 10, 11, 12, and 13, starting with mean magnitudes, pulsation periods, time of brightness maxima, and pulsation amplitude. The following two columns list the RR~Lyrae pulsation type and its conditional probability. The last columns flag stars that were associated with the Orphan stream by \citet[marked with K19,][]{Koposov2019Orphan} as parent population. The asterisk at \texttt{bestObjID} indicates a star that was not used as reference sample in Sec.~\ref{sec:stableOrphan}.}
\label{tab:Orphan-RR}
\begin{tabular}{lcccccccccccccccc}
\hline
\texttt{bestObjID} (SDSS) & \gaia~EDR3 ID & $\alpha$ & $\delta$ & $d$ & $\sigma_{d}$ & $v_{\rm sys}$ & $\sigma_{v_{\rm sys}}$ & $V_{\rm CSS}$ & $\sigma_{V_{\rm CSS}}$ & $P$ & $M_{0}$ & Amp$_{V_{\rm CSS}}$ & Type & [Fe/H] & $p(A|B)$ & Note \\ 
 & & [deg] & [deg] & [kpc] & [kpc] & [km\,s$^{-1}$] & [km\,s$^{-1}$] & [mag] & [mag] & [day] & [day] & [mag] & & [dex] & & \\ \hline
1237660635454701712 & 801408324401633664 & 146.05782 & 40.22071 & 39.7 & 2.2 & 144.742 & 18.081 & 18.453 & 0.135 & 0.711533 & 54588.22739 & 0.57071 & RRab & -1.680 & 0.272 & K19 \\ 
1237660635453718722 & 812926670775689984 & 143.48258 & 39.13402 & 42.5 & 2.3 & 172.430 & 15.680 & 18.582 & 0.145 & 0.527852 & 55198.25415 & 0.64537 & RRab & -1.880 & 0.111 & K19 \\ 
1237667734496018571 & 625033259008713344 & 153.80169 & 19.05096 & 26.3 & 1.5 & 199.518 & 6.169 & 17.703 & 0.100 & 0.400190 & 54769.47305 & 0.44528 & RRc & -2.030 & 0.915 & --- \\ 
1237657770706600085 & 1011841380940809344 & 140.40968 & 48.01452 & 45.4 & 2.5 & 109.058 & 1.139 & 18.922 & 0.158 & 0.367648 & 56402.27718 & 0.37343 & RRc & -2.030 & 0.592 & --- \\ 
1237667782285131881 & 625042020741726976 & 153.99639 & 19.22272 & 25.8 & 1.4 & 214.922 & 7.709 & 17.667 & 0.098 & 0.645172 & 55563.40418 & 0.60377 & RRab & -1.720 & 0.799 & --- \\ 
1237668290157281403 & 623982645584012928 & 154.82491 & 18.22602 & 28.0 & 1.6 & 194.632 & 10.047 & 17.870 & 0.108 & 0.578450 & 54628.16950 & 0.77584 & RRab & -1.670 & 0.239 & K19 \\ 
1237657606967459944 & 1011263007760611456 & 139.35631 & 46.72456 & 42.3 & 2.3 & 86.520 & 12.399 & 18.762 & 0.155 & 0.388203 & 54862.20461 & 0.35120 & RRc & -1.870 & 0.194 & --- \\ 
1237657773935624421 & 814812268794932608 & 144.29504 & 43.42943 & 41.5 & 2.3 & 140.578 & 6.938 & 18.824 & 0.156 & 0.366009 & 54535.33735 & 0.37174 & RRc & -1.380 & 0.358 & K19 \\ 
1237658203425341674 & 813632316722202112 & 145.61867 & 41.56253 & 42.1 & 2.3 & 157.187 & 17.406 & 18.489 & 0.134 & 0.604208 & 55505.50362 & 0.58930 & RRab & -2.190 & 0.188 & K19 \\ 
1237661851456962762 & 800283700102935808 & 147.37900 & 38.73692 & 37.6 & 2.1 & 166.666 & 6.772 & 18.135 & 0.117 & 0.286424 & 55212.30609 & 0.20096 & RRc & -1.910 & 0.105 & K19 \\ 
1237664870825918615 & 793317812902061568 & 147.81260 & 32.49737 & 39.3 & 2.2 & 171.941 & 13.919 & 18.533 & 0.134 & 0.552830 & 54035.50252 & 0.65775 & RRab & -1.290 & 0.195 & --- \\ 
1237665099002937435 & 744466232107315712 & 148.36049 & 30.02346 & 38.0 & 2.1 & 208.007 & 4.582 & 18.392 & 0.128 & 0.591062 & 53677.59556 & 0.40989 & RRab & -2.150 & 0.265 & K19 \\ 
1237665129604317276 & 744807802266002432 & 148.91221 & 30.42627 & 34.8 & 1.9 & 197.726 & 35.431 & 18.390 & 0.129 & 0.360622 & 54574.20862 & 0.37578 & RRc & -1.450 & 0.756 & --- \\ 
1237660634916913359$^{\ast}$ & 800895883264108416 & 143.91322 & 38.85322 & 43.1 & 2.4 & 53.148 & 12.118 & 18.766 & 0.150 & 0.504141 & 55566.42701 & 0.58476 & RRab & -1.780 & 0.366 & K19 \\ 
1237660763234107516 & 799585024885253632 & 147.19268 & 37.13167 & 40.3 & 2.2 & 166.001 & 4.421 & 18.579 & 0.140 & 0.624428 & 53902.06206 & 0.62468 & RRab & -1.870 & 0.068 & K19 \\ 
1237661850382696669 & 799823623206083328 & 146.44757 & 37.55324 & 39.9 & 2.2 & 140.716 & 10.239 & 18.329 & 0.130 & 0.624027 & 56313.47731 & 1.02146 & RRab & -1.820 & 0.524 & K19 \\ 
1237661139030966463 & 799463292628940672 & 146.00854 & 36.26583 & 40.5 & 2.3 & 156.131 & 2.478 & 18.546 & 0.137 & 0.594447 & 54532.36231 & 0.76866 & RRab & -1.880 & 0.283 & K19 \\ 
1237657874867421307 & 814635483643723520 & 144.27165 & 42.60335 & 43.3 & 2.4 & 160.573 & 12.657 & 18.579 & 0.136 & 0.567186 & 54789.53586 & 0.65984 & RRab & -1.650 & 0.461 & --- \\ 
1237667252929036352 & 738597146412411904 & 151.89251 & 24.83150 & 31.0 & 1.8 & 180.319 & 22.057 & 17.944 & 0.114 & 0.620870 & 54382.54030 & 1.03188 & RRab & -2.150 & 0.764 & K19 \\ 
1237657771780866136 & 1018131343366018560 & 141.13135 & 49.38273 & 46.9 & 2.7 & 111.560 & 8.632 & 19.028 & 0.161 & 0.342678 & 55119.41420 & 0.52122 & RRc & -1.500 & 0.165 & --- \\ \hline
\end{tabular}
\end{table}
\end{landscape}

\begin{landscape}
\begin{table}
\caption{List of nonvariable stars associated with the Orphan stellar stream based on our RR~Lyrae sample. The first two columns list the identifiers from the SDSS and \gaia~EDR3, the following two columns the objects equatorial coordinates. Columns 5 and 6, contain the line-of-sight velocities determined by the SSPP pipeline, and the subsequent two columns provide their $g$-band magnitudes together with their uncertainties. Columns 9, 10, 11, 12, 13, and 14 list the stellar parameters derived by the SSPP pipeline. The last column represents the conditional probability for the individual star. The asterisk at \texttt{bestObjID} marks stars that are classified as RRd type stars or their classification in RR~Lyrae subtypes is uncertain.} 
\label{tab:Orphan-Stab}
\begin{tabular}{lcccccccccccccc}
\hline
\texttt{bestObjID} (SDSS) & \gaia~EDR3 ID & $\alpha$ & $\delta$ & \texttt{RV\_ADOP} & \texttt{RV\_ADOP\_UNC} & $g$ & $\sigma_{g}$ & $T_{\text{eff}}$ & $\sigma_{T_{\text{eff}}}$ & [Fe/H] & $\sigma_{\text{[Fe/H]}}$ & log\,$g$ & $\sigma_{\text{log}\,g}$ & $p(A|B)$ \\
 & & [deg] & [deg] & [km\,s$^{-1}$] & [km\,s$^{-1}$] & [mag] & [mag] & [K] & [K] & [dex] & [dex] & [dex] & [dex] & \\ \hline
1237667537471144142 & 628696866112455168 & 151.60904 & 21.04929 & 221.210 & 14.141 & 19.493 & 0.025 & 8311 & 253 & -1.86 & 0.38 & 3.69 & 0.10 & 0.137 \\ 
1237667211053498537 & 738657310314238720 & 152.20035 & 25.46991 & 198.841 & 3.985 & 17.955 & 0.019 & 8349 & 68 & -1.64 & 0.08 & 3.33 & 0.39 & 0.402 \\ 
1237667211590566073 & 738839309553409536 & 152.65131 & 26.09226 & 185.870 & 9.608 & 20.035 & 0.025 & 5162 & 216 & -2.04 & 0.09 & 2.43 & 0.54 & 0.242 \\ 
1237667549803446363 & 628835095339982976 & 153.21074 & 21.01953 & 197.102 & 2.124 & 17.954 & 0.022 & 5046 & 11 & -2.03 & 0.08 & 1.92 & 0.17 & 0.893 \\ 
1237667430635143257 & 630353112875731840 & 151.01106 & 23.71998 & 201.503 & 2.698 & 16.792 & 0.018 & 6107 & 102 & -1.91 & 0.05 & 2.17 & 0.29 & 0.285 \\ 
1237667736106369165 & 625374592944708864 & 152.97943 & 20.03164 & 223.173 & 5.468 & 18.203 & 0.017 & 5178 & 25 & -2.36 & 0.08 & 2.05 & 0.14 & 0.539 \\ 
1237667551413796866 & 629200481092312064 & 152.26792 & 22.22917 & 199.110 & 1.543 & 16.744 & 0.026 & 4681 & 105 & -2.02 & 0.01 & 1.50 & 0.08 & 0.830 \\ 
1237667537471930559 & 628871997698600704 & 153.51826 & 21.40729 & 193.410 & 3.993 & 18.198 & 0.021 & 5182 & 55 & -2.27 & 0.06 & 2.01 & 0.07 & 0.824 \\ 
1237667252929167431 & 726590372062626944 & 152.22576 & 24.81725 & 200.607 & 3.832 & 17.953 & 0.026 & 8139 & 139 & -2.24 & 0.10 & 3.21 & 0.20 & 0.490 \\ 
1237667253466103892 & 738656996781344128 & 152.20437 & 25.42895 & 209.124 & 5.578 & 17.728 & 0.019 & 8359 & 47 & -1.97 & 0.08 & 3.41 & 0.35 & 0.760 \\ 
1237667210516562002 & 738630681517008000 & 152.19512 & 25.11815 & 197.252 & 4.560 & 17.599 & 0.017 & 5184 & 50 & -2.18 & 0.03 & 2.42 & 0.21 & 0.249 \\ 
1237667736106303744 & 625388813581465856 & 152.80820 & 20.13625 & 207.441 & 8.433 & 18.030 & 0.025 & 8733 & 295 & -1.90 & 0.08 & 3.09 & 0.50 & 0.645 \\ 
1237660343936090311 & 812965188042058752 & 143.88747 & 39.66263 & 130.608 & 11.815 & 18.741 & 0.024 & 8381 & 214 & -1.24 & 0.14 & 3.19 & 0.65 & 0.181 \\ 
1237667430635536640 & 630417842327675776 & 152.07167 & 23.92788 & 183.996 & 4.831 & 18.144 & 0.018 & 7219 & 113 & -1.70 & 0.10 & 3.22 & 0.25 & 0.885 \\ 
1237657776082518150 & 817873957704595328 & 141.36492 & 44.12217 & 113.810 & 3.612 & 18.336 & 0.014 & 5042 & 44 & -2.09 & 0.07 & 1.46 & 0.13 & 0.124 \\ 
1237667735570088150 & 625510618853633152 & 154.46992 & 19.93856 & 207.081 & 5.169 & 18.514 & 0.022 & 5324 & 44 & -1.94 & 0.16 & 2.40 & 0.09 & 0.658 \\ 
1237661383846920453 & 796532505729010048 & 147.39971 & 36.55098 & 151.613 & 3.837 & 18.422 & 0.017 & 5039 & 44 & -2.11 & 0.04 & 2.11 & 0.08 & 0.173 \\ 
1237664667895398511$^{\ast}$ & 793633269661481344 & 147.67491 & 33.13807 & 166.882 & 8.848 & 18.772 & 0.026 & 6795 & 75 & -1.50 & 0.06 & 2.73 & 0.39 & 0.761 \\ 
1237668289083736145 & 3890404706979164928 & 155.38993 & 17.49104 & 212.624 & 5.389 & 18.767 & 0.027 & 5348 & 53 & -2.64 & 0.01 & 2.21 & 0.32 & 0.965 \\ 
1237660764307128401 & 799763802901092864 & 144.96541 & 37.18751 & 151.803 & 1.014 & 16.776 & -9999 & 4499 & 409 & -1.84 & 0.05 & 0.81 & 0.32 & 0.535 \\ 
1237657628979953819 & 815043750351075328 & 143.73901 & 44.08255 & 119.774 & 5.052 & 18.225 & 0.013 & 5233 & 64 & -2.32 & 0.03 & 1.67 & 0.43 & 0.206 \\ 
1237657606967459944 & 1011263007760611456 & 139.35631 & 46.72456 & 86.322 & 9.121 & 19.039 & 0.019 & 6978 & 152 & -1.92 & 0.04 & 3.27 & 0.64 & 0.922 \\ 
1237657773935624352 & 814827352720083968 & 144.07772 & 43.51372 & 120.661 & 2.524 & 18.215 & 0.027 & 5009 & 127 & -1.81 & 0.04 & 1.99 & 0.24 & 0.108 \\ 
1237658205035364447 & 814233646503353984 & 143.00382 & 42.01405 & 130.019 & 3.959 & 18.514 & 0.015 & 5220 & 243 & -1.58 & 0.15 & 1.99 & 0.55 & 0.533 \\ 
1237667733959082068 & 624157566716484096 & 153.72088 & 18.44767 & 214.915 & 4.277 & 18.342 & 0.020 & 5167 & 22 & -2.03 & 0.04 & 2.26 & 0.12 & 0.371 \\ 
1237667549266509948 & 625435577185459968 & 153.25189 & 20.63283 & 186.624 & 11.460 & 19.154 & 0.018 & 8890 & 370 & -2.01 & 0.12 & 3.57 & 0.19 & 0.928 \\  \hline
 & &  & & & & & & & & & & & & \textit{Continued on next page} \\
\end{tabular}
\end{table}
\end{landscape}

\begin{landscape}  
\begin{table}
\caption{\textit{Continued from previous page}}
\begin{tabular}{lcccccccccccccc}
\hline
\texttt{bestObjID} (SDSS) & \gaia~EDR3 ID & $\alpha$ & $\delta$ & \texttt{RV\_ADOP} & \texttt{RV\_ADOP\_UNC} & $g$ & $\sigma_{g}$ & $T_{\text{eff}}$ & $\sigma_{T_{\text{eff}}}$ & [Fe/H] & $\sigma_{\text{[Fe/H]}}$ & log\,$g$ & $\sigma_{\text{log}\,g}$ & $p(A|B)$ \\
 & & [deg] & [deg] & [km\,s$^{-1}$] & [km\,s$^{-1}$] & [mag] & [mag] & [K] & [K] & [dex] & [dex] & [dex] & [dex] & \\ \hline
1237657628979757237 & 815002102051985536 & 143.05483 & 43.85158 & 105.393 & 7.243 & 18.967 & 0.015 & 7892 & 106 & -1.83 & 0.10 & 3.46 & 0.25 & 0.355 \\ 
1237667254540173418 & 738864082924453888 & 152.62750 & 26.39346 & 194.872 & 4.968 & 19.065 & 0.020 & 5414 & 35 & -2.33 & 0.05 & 2.60 & 0.14 & 0.501 \\ 
1237667253466431572 & 726732209062781056 & 152.88857 & 25.52055 & 195.383 & 6.763 & 18.022 & 0.029 & 8132 & 61 & -1.95 & 0.03 & 3.33 & 0.20 & 0.509 \\ 
1237661384383266935 & 799481881247354496 & 145.68937 & 36.40120 & 150.277 & 2.854 & 17.956 & 0.018 & 4970 & 57 & -2.10 & 0.06 & 1.23 & 0.32 & 0.303 \\ 
1237664338242896034$^{\ast}$ & 745248362831928064 & 149.35785 & 32.02152 & 164.780 & 7.265 & 18.302 & 0.023 & 6831 & 120 & -2.10 & 0.09 & 3.92 & 0.46 & 0.577 \\ 
1237661383846461589 & 799409377903432192 & 146.20282 & 36.03515 & 156.220 & 9.043 & 18.651 & 0.018 & 8268 & 225 & -2.68 & 0.14 & 3.03 & 0.11 & 0.322 \\ 
1237657628442427527 & 814555842066743680 & 142.28329 & 42.77482 & 120.226 & 3.019 & 18.159 & 0.014 & 4819 & 145 & -1.93 & 0.08 & 1.29 & 0.27 & 0.819 \\ 
1237660764307128471 & 799764417079227776 & 144.92913 & 37.18107 & 147.710 & 12.770 & 18.565 & 0.014 & 7988 & 179 & -1.56 & 0.09 & 3.15 & 0.71 & 0.308 \\ 
1237667429562122398 & 630132244182398336 & 153.00681 & 23.37064 & 196.173 & 5.322 & 18.106 & 0.024 & 6179 & 158 & -1.65 & 0.04 & 1.61 & 0.38 & 0.124 \\ 
1237660763234107445 & 799585750736939136 & 147.17276 & 37.13210 & 156.173 & 1.953 & 17.469 & 0.026 & 4823 & 103 & -2.38 & 0.06 & 1.71 & 0.09 & 0.580 \\ 
1237661851455848669$^{\ast}$ & 800538438905600640 & 144.52811 & 37.58107 & 119.584 & 7.176 & 18.934 & 0.026 & 7130 & 96 & -2.85 & 0.25 & 2.69 & 0.47 & 0.055 \\ 
1237657874330484837 & 813866547058758400 & 144.33302 & 42.31582 & 122.986 & 8.127 & 18.775 & 0.018 & 7666 & 50 & -2.30 & 0.10 & 3.68 & 0.43 & 0.558 \\ 
1237662224591356037 & 794981472779226880 & 148.48536 & 33.61997 & 158.403 & 10.066 & 18.455 & 0.013 & 7589 & 103 & -1.88 & 0.06 & 3.92 & 0.34 & 0.677 \\ 
1237668288546865284 & 3890324339551568896 & 155.43794 & 16.96701 & 205.456 & 3.210 & 17.861 & 0.019 & 8204 & 31 & -1.71 & 0.05 & 3.30 & 0.32 & 0.742 \\ 
1237660763233976439 & 799560973068347520 & 146.86139 & 36.97398 & 149.349 & 6.724 & 18.519 & 0.020 & 8904 & 412 & -1.10 & 0.53 & 3.49 & 0.46 & 0.078 \\ 
1237660764844130395 & 800525317782751104 & 145.03387 & 37.48114 & 147.770 & 8.381 & 18.710 & 0.021 & 8643 & 218 & -2.27 & 0.10 & 3.36 & 0.23 & 0.522 \\ 
1237660343936483487 & 801059714792493312 & 145.00683 & 39.96981 & 124.638 & 5.739 & 19.099 & 0.022 & 5738 & 112 & -1.97 & 0.05 & 2.61 & 0.26 & 0.400 \\ 
1237670965928788035$^{\ast}$ & 623884479811567232 & 154.79668 & 18.09111 & 201.940 & 3.353 & 17.347 & 0.053 & 7763 & 122 & -1.14 & 0.38 & 3.10 & 0.21 & 0.664 \\ 
1237661850382696658 & 799820595251815552 & 146.47956 & 37.43788 & 142.935 & 3.601 & 18.020 & 0.015 & 4895 & 108 & -2.45 & 0.05 & 1.27 & 0.16 & 0.156 \\ 
1237667255076454406 & 738998051544247808 & 151.09596 & 26.26995 & 186.114 & 2.682 & 17.149 & 0.022 & 4924 & 66 & -2.44 & 0.07 & 1.77 & 0.09 & 0.068 \\ 
1237660763769995449 & 798207787789021056 & 144.56194 & 36.46507 & 161.170 & 3.906 & 18.441 & 0.020 & 5135 & 43 & -2.18 & 0.07 & 1.68 & 0.21 & 0.101 \\ 
1237668290157543482 & 623997244177536512 & 155.49606 & 18.33054 & 211.947 & 3.323 & 17.821 & 0.022 & 8103 & 18 & -1.81 & 0.02 & 3.47 & 0.08 & 0.391 \\ 
1237667211590172695 & 738914454301088896 & 151.55458 & 25.85548 & 196.119 & 2.262 & 17.157 & 0.014 & 4889 & 64 & -2.06 & 0.05 & 1.59 & 0.12 & 0.241 \\ 
1237670964318371898 & 3890296095846572160 & 155.49308 & 16.83852 & 217.476 & 1.614 & 17.537 & 0.026 & 5071 & 76 & -2.00 & 0.04 & 1.79 & 0.11 & 0.555 \\ 
1237660635454505093 & 801391389347743104 & 145.43703 & 40.03586 & 137.319 & 2.639 & 17.805 & 0.019 & 4909 & 33 & -2.20 & 0.03 & 1.58 & 0.09 & 0.430 \\ 
1237660962942943451 & 801079020669410048 & 146.22752 & 38.75768 & 146.472 & 7.459 & 18.594 & 0.017 & 7804 & 188 & -1.75 & 0.14 & 3.29 & 0.33 & 0.679 \\ 
1237667783895351306 & 625392769246309504 & 152.72107 & 20.20343 & 204.944 & 1.443 & 16.662 & 0.022 & 4687 & 156 & -1.94 & 0.05 & 1.34 & 0.07 & 0.870 \\ 
1237661139031425067 & 796535460666541696 & 147.24772 & 36.61864 & 155.236 & 5.931 & 18.332 & 0.028 & 6772 & 228 & -1.25 & 0.16 & 2.46 & 0.57 & 0.353 \\ \hline
\end{tabular}
\end{table}
\end{landscape}
\end{appendix}


\end{document}